\makeatletter
\let\orilabel\label
\documentclass[aps,prd,reprint,a4paper,amsmath,amsfonts,amssymb,nofootinbib]{revtex4-2}
\let\label\orilabel
\let\ltx@label\label
\makeatother
\usepackage[utf8]{inputenc}
\usepackage[colorlinks=true,urlcolor=blue,anchorcolor=blue,citecolor=blue,filecolor=blue,linkcolor=blue,menucolor=blue,linktocpage=true,bookmarksopen,pdfa=true,unicode]{hyperref}

\usepackage{mathtools}
\usepackage{lmodern}
\usepackage[T1]{fontenc}
\usepackage[bb=stix]{mathalpha}
\DeclareMathAlphabet{\mathbfi}{OML}{cmm}{b}{it}

\let\originalleft\left
\let\originalright\right
\renewcommand{\left}{\mathopen{}\mathclose\bgroup\originalleft}
\renewcommand{\right}{\aftergroup\egroup\originalright}

\makeatletter
\newenvironment{equations}[1][]{\subequations\ifx\relax#1\relax\else\label{#1}\fi\align\ignorespaces}{\endalign\ignorespacesafterend\endsubequations}
\def\@spliteq#1{\begin{equation}\begin{split}#1\end{split}\end{equation}}
\def\splitequation{\collect@body\@spliteq}

\makeatother

\newcommand{\eqend}[1]{\,\mathrm{#1}}
\newcommand{\bra}[1]{{\left\langle{#1}\right\rvert}}
\newcommand{\ket}[1]{{\left\lvert{#1}\right\rangle}}

\newcommand{\abs}[1]{{\left\lvert{#1}\right\rvert}}
\newcommand{\laplace}{\bigtriangleup}

\newcommand{\bessel}[3]{\mathrm{#1}_{#2}\left(#3\right)}

\newcommand{\hypergeom}[2]{\,{}_{#1}\mathrm{F}_{#2}}
\newcommand{\sgn}{\mathop{\mathrm{sgn}}}

\newcommand{\total}{\mathop{}\!\mathrm{d}}
\renewcommand{\vec}[1]{{\ifnum9<1#1\mathbf{#1}\else\ifcat\noexpand#1\relax\boldsymbol{#1}\else\mathbfi{#1}\fi\fi}}
\newcommand{\mathi}{\mathrm{i}}
\newcommand{\mathe}{\mathrm{e}}

\newcommand{\dS}{\mathbb{dS}}
\newcommand{\tr}{\operatorname{tr}}
\newcommand{\supp}{\operatorname{supp}}
\newcommand{\symp}[1]{{\sigma\left({#1}\right)}}

\begin{document}

\title{Relative Entropy in de~Sitter is a Noether Charge}

\author{Markus B. Fröb}
\email{mfroeb@itp.uni-leipzig.de}
\author{Albert Much}
\email{much@itp.uni-leipzig.de}
\affiliation{Institut f{\"u}r Theoretische Physik, Universit{\"a}t Leipzig, Br{\"u}derstra{\ss}e 16, 04103 Leipzig, Germany}
\author{Kyriakos Papadopoulos}
\email[Corresponding author: ]{kyriakos@sci.kuniv.edu.kw}
\affiliation{Department of Mathematics, Faculty of Science, Kuwait University, Safat 13060, Kuwait}

\begin{abstract}
We compute the relative entropy between the vacuum and a coherent state for a massive scalar field in de~Sitter spacetime, using Tomita--Takesaki modular theory and the Araki--Uhlmann formula for the relative entropy. Embedding de~Sitter spacetime as a hyperboloid in the ambient Minkowski space, we can restrict the Minkowski wedge and the corresponding modular operator to de~Sitter, and we verify that this construction gives the correct modular flow. We check that the relative entropy is positive and jointly convex, relate it to the Noether charge of translations along the trajectories of the modular flow, and determine the local temperature as seen by an observer that moves along these trajectories.
\end{abstract}

\maketitle

\section{Introduction}
\label{sec:intro}

Entanglement entropy, computed as the von Neumann entropy $\mathcal{S}_\text{vN}(\rho) = - \tr( \rho \ln \rho )$ of the reduced density matrix $\rho_R$ of the degrees of freedom localized in some region or subsystem $R$, is an important measure of entanglement in statistical physics. However, its application to quantum field theory shows some problems. First, it is a divergent quantity in the continuum (see for example Ref.~\cite{marolfwall2016} and references therein): the entanglement entropy $\mathcal{S}$ of a region in an $n$-dimensional spacetime computed with an UV cutoff $\epsilon$ behaves like $\mathcal{S} \sim A \, \epsilon^{2-n}$ in the limit $\epsilon \to 0$, where $A$ is the $(n-2)$-dimensional area of the region's boundary. Second, the proportionality coefficient depends on the number of fields in the theory and the details of their interaction. While this formula correctly reproduces the area dependence of the famous Bekenstein--Hawking formula for the entropy of a black hole horizon~\cite{bekenstein1973,bekenstein1974,hawking1974,hawking1975}, the latter is not only finite (with the UV cutoff replaced by the Planck length $\ell_\text{Pl}$), but also universal with a model-independent coefficient of proportionality equal to $1/4$.

To be able to obtain the Bekenstein--Hawking formula from entropy considerations in quantum field theory, one therefore has to study finite quantities such as the relative entropy
\begin{equation}
\label{eq:relative_entropy}
\mathcal{S}(\rho\Vert\sigma) = \tr\left( \rho \ln \rho - \rho \ln \sigma \right) \eqend{,}
\end{equation}
which compares two different density matrices $\rho$ and $\sigma$. In particular, if $\rho$ and $\sigma$ are the reduced density matrices of two different states, one obtains the relative entanglement entropy that compares the entanglement of these states. Since the divergences in the entanglement entropy essentially result from the high-frequency modes which are common to all states (including the vacuum, which results in the Reeh--Schlieder theorem~\cite{reehschlieder1961}), the relative entanglement entropy can be finite also in quantum field theory in the limit of a vanishing UV cutoff $\epsilon$. That this is indeed so is shown by the Araki--Uhlmann formula~\cite{araki1975,araki1976,uhlmann1977}
\begin{equation}
\label{eq:araki_uhlmann}
\mathcal{S}(\Psi\Vert\Phi) = - \bra{\Psi} \ln \Delta_{\Psi\vert\Phi} \ket{\Psi} \eqend{,}
\end{equation}
which relates the relative entropy between two states $\ket{\Psi}$ and $\ket{\Phi}$ to the expectation value of the relative modular Hamiltonian $\ln \Delta_{\Psi\vert\Phi}$ associated to these states and a given von Neumann algebra $\mathcal{A}$. This formula was obtained in the framework of Tomita--Takesaki modular theory~\cite{tomita1967,takesaki1970}, where the von Neumann algebra $\mathcal{A}$ can be taken to be the algebra of fields in a certain region (such as a wedge or a double cone), and where the states $\ket{\Psi}$ and $\ket{\Phi}$ must be cyclic and separating for $\mathcal{A}$.

In quantum mechanics, the relative modular Hamiltonian can be written in terms of the reduced density matrices $\rho_\Psi$ and $\rho_\Phi$ associated to the states $\ket{\Psi}$ and $\ket{\Phi}$. In fact, on the tensor product Hilbert space describing the bipartite quantum system (of the region of interest and its complement) it has the very simple expression~\cite{witten2018}
\begin{equation}
\ln \Delta_{\Psi\vert\Phi} = \ln\left( \rho_\Phi^{-1} \otimes \rho_\Psi \right) \eqend{,}
\end{equation}
such that the Araki--Uhlmann formula~\eqref{eq:araki_uhlmann} reduces to
\begin{equation}
\mathcal{S}(\Psi\Vert\Phi) = - \tr\left[ \rho_\Psi \ln\left( \rho_\Phi^{-1} \otimes \rho_\Psi \right) \right] = \mathcal{S}(\rho_\Psi\Vert\rho_\Phi) \eqend{,}
\end{equation}
the correct expression~\eqref{eq:relative_entropy} for the relative entropy in terms of density matrices. In quantum field theory, the determination of the relative modular Hamiltonian is more difficult. One important subcase is when the states $\ket{\Psi}$ and $\ket{\Phi}$ are obtained as coherent excitations of another state $\ket{\Omega}$, for which the modular Hamiltonian $\ln \Delta_\Omega$ is known. In this case, one has $\ket{\Psi} = U \ket{\Omega}$ and $\ket{\Phi} = V \ket{\Omega}$ with unitary operators $U, V \in \mathcal{A}$, and it is easy to show~\cite{witten2018,casinigrillopontello2019,lashkariliurajagopal2021} that $\ln \Delta_{\Psi\vert\Phi} = V \ln \Delta_\Omega V^\dagger$ and consequently
\begin{equation}
\label{eq:araki_uhlmann_coherent}
\mathcal{S}(\Psi\Vert\Phi) = - \bra{\Omega} U^\dagger V \ln \Delta_\Omega V^\dagger U \ket{\Omega} \eqend{.}
\end{equation}
In this way and using the known result for the modular Hamiltonian in wedges (the Bisognano--Wichmann theorem~\cite{bisognanowichmann1975,bisognanowichmann1976}), the relative entropy for coherent excitations of the vacuum of a free massive scalar field, restricted to a wedge in Minkowski spacetime, was computed by various authors~\cite{casinigrillopontello2019,longo2019}.

For the exterior of a Schwarzschild black hole, the modular Hamiltonian of free scalar fields in the Hartle--Hawking state is also known~\cite{kay1985a,kay1985b,sewell1982,kaywald1991}, and in fact proportional to the generator of time translations; if one restricts to the black hole horizon or null infinity, these become rescalings of the corresponding null coordinates~\cite{brunettiguidolongo2002}. Using this fact, the Araki--Uhlmann formula for the relative entropy of coherent states~\eqref{eq:araki_uhlmann_coherent} and the Raychaudhuri equation, Hollands and Ishibashi~\cite{hollandsishibashi2019} have shown that the variation of the sum of relative entropy and one quarter of the horizon area is exactly given by the flux at future null infinity, in line with the Bekenstein--Hawking formula for the entropy of a black hole horizon. These results have been further generalized to apparent horizons by D'Angelo~\cite{dangelo2020,dangelo2021}. On the other hand, Iyer and Wald~\cite{wald1993,iyerwald1994} have shown that black hole entropy can be alternatively defined as the Noether charge of diffeomorphisms associated with the Killing vector field of the black hole horizon, integrated over it. However, a connection between the two approaches has not been made.

In this article, we make a related but different connection. First, we generalize the results of~\cite{casinigrillopontello2019,longo2019} to wedges in the de~Sitter spacetime, and compute the relative entropy for a coherent excitation of the de~Sitter vacuum of a free massive scalar field. We confirm that the relative entropy is positive and convex as required, and finally relate it to the Noether charge of translations along the Killing vector that is tangent to the modular flow. De~Sitter spacetime is of course important as a model of both the primordial inflationary phase of the universe and the current exponential expansion~\cite{riesetal1998,perlmutteretal1999,planck2018}. At the same time, it can be obtained from an embedding in Minkowski spacetime of one higher dimension and is a maximally symmetric solution of the Einstein equations with cosmological constant, therefore one of the simplest curved spacetimes. Among others, this also manifests in the fact that the modular Hamiltonian for wedges in de~Sitter spacetime is known as well~\cite{brosepsteinmoschella1998,borchersbuchholz1999,buchholzdreyerflorigsummers2000}, and in fact coincides with the restriction of the modular Hamiltonian of the wedge in the embedding Minkowski spacetime to the de~Sitter hyperboloid.

The remainder of this article is structured as follows: In Sec.~\ref{sec:desitter}, we review the embedding of de~Sitter space, the restriction of the Minkowski Killing vectors to the de~Sitter hyperboloid, and the wedges in both Minkowski and de~Sitter spacetime. Section~\ref{sec:scalarmodular} is devoted to the algebra of the free scalar field and modular theory, and we determine explicitly the modular flow and the Kubo--Martin--Schwinger (KMS) property for the de~Sitter vacuum state restricted to a wedge, following in part Ref.~\cite{borchersbuchholz1999}. In Sec.~\ref{sec:entropy}, we then determine the relative entropy between the de~Sitter vacuum and a coherent state using the Araki--Uhlmann formula~\eqref{eq:araki_uhlmann_coherent} and verify its positivity and joint convexity. Section~\ref{sec:noether} is concerned with the relation of the relative entropy to a Noether charge, and its thermodynamic interpretation. We give an outlook on future extensions of this work in Sec.~\ref{sec:discussion}, and leave the details of the covariant canonical quantization to Appendix~\ref{sec:app_quantization}, some further relations that are satisfied by the commutator function to Appendices~\ref{sec:app_symplectic} and~\ref{sec:app_commutator}, and details of the computation for a tilted Cauchy surface to Appendix~\ref{sec:app_cauchy}.

\paragraph*{Conventions:} We take mostly plus metric signature, and choose $R_{\mu\nu} = (n-1) H^2 g_{\mu\nu}$ for $n$-dimensional de~Sitter spacetime of radius $H^{-1}$. All formulas for special functions were taken from Ref.~\cite{dlmf}. Greek indices $\mu,\nu,\dots \in \{0,\ldots,n-1\}$ range over space and time, while lowercase Latin indices $i,j,\dots \in \{1,\ldots,n-1\}$ denote purely spatial components, and uppercase Latin indices $A,B,\dots \in \{0,\ldots,n\}$ refer to the embedding (or ambient) space. We set $\hbar = c = 1$.

\section{de~Sitter spacetime}
\label{sec:desitter}

As stated in the introduction, $n$-dimensional de~Sitter spacetime $\dS_n$ can be obtained from an embedding in an $(n+1)$-dimensional Minkowski spacetime $\mathbb{R}^{n,1}$ (called ambient space). Choosing Cartesian coordinates $X^A$ for the ambient space, $\dS_n$ is the submanifold of points satisfying $X^A X_A = H^{-2}$ with a constant $H$, the Hubble rate or inverse de~Sitter radius. The relevant part of $\dS_n$ is the so-called expanding Poincar{\'e} patch (with $X^0 - X^n > 0$), which can be parametrized by coordinates $x^0 = t \in \mathbb{R}$ and $\vec{x} \in \mathbb{R}^{n-1}$ according to
\begin{equations}[eq:poincarepatch]
X^0 &= \frac{1}{H} \sinh(H t) + \frac{H}{2} \mathe^{H t} \vec{x}^2 \eqend{,} \\
X^i &= \mathe^{H t} \vec{x}^i \eqend{,} \\
X^n &= - \frac{1}{H} \cosh(H t) + \frac{H}{2} \mathe^{H t} \vec{x}^2 \eqend{.}
\end{equations}
It is easy to verify that these satisfy the hyperboloid condition $X^A X_A = H^{-2}$, and the induced metric of the Poincar{\'e} patch reads
\begin{equation}
\label{eq:poincare_metric}
\total s^2 = \eta_{AB} \total X^A \total X^B = - \total t^2 + \mathe^{2 H t} \total \vec{x}^2 \eqend{.}
\end{equation}
The causal structure of the de~Sitter spacetime is inherited through the embeddinging, and the distance between points can be determined from the ambient space as well. It is characterized by the invariant
\begin{splitequation}
\label{eq:z_embedding}
Z(x,x') &= H^2 \eta_{AB} X^A(x) X^B(x') \\
&= 1 - \frac{H^2}{2} \big[ X_A(x) - X_A(x') \big] \big[ X^A(x) - X^A(x') \big] \eqend{,} \raisetag{3.6em}
\end{splitequation}
which in terms of the Poincar{\'e} patch coordinates~\eqref{eq:poincarepatch} reads
\begin{equation}
\label{eq:z_def}
Z(x,x') = \cosh[ H (t-t') ] - \frac{H^2}{2} \mathe^{H (t+t')} (\vec{x}-\vec{x}')^2 \eqend{.}
\end{equation}
From Eq.~\eqref{eq:z_embedding}, we see that spacelike separated points have $Z(x,x') < 1$, lightlike separation renders $Z(x,x') = 1$, and if $Z(x,x') > 1$ the points $x$ and $x'$ are timelike separated.

The ambient space has $(n+1)(n+2)/2$ Killing vectors, of which $n+1$ generate translations and $n(n+1)/2$ generate rotations and boosts. While the translation generators $T_A = \partial_A$ do not leave the hyperboloid invariant, the rotation and boost generators $M_{AB} = X_A \partial_B - X_B \partial_A$ transform the hyperboloid into itself and hence descend to Killing vectors in $\dS_n$. Expressing the Poincar{\'e} patch coordinates~\eqref{eq:poincarepatch} using the embedding coordinates, we obtain
\begin{equations}
t &= H^{-1} \ln [ H ( X^0 - X^n ) ] \eqend{,} \\
\vec{x}^i &= \frac{X^i}{H ( X^0 - X^n )} \eqend{,}
\end{equations}
which allows us to compute the explicit expression of the rotation and boost generators in terms of $t$ and $\vec{x}$. Concretely, we find
\begin{equation}
\label{eq:minkowski_killing}
\begin{aligned}
M_{0i} &= - \frac{H}{2} K_i - \frac{1}{2 H} P_i \eqend{,} \quad && M_{ij} = L_{ij} \eqend{,} \\
M_{in} &= - \frac{H}{2} K_i + \frac{1}{2 H} P_i \eqend{,} \quad && M_{0n} = - D
\end{aligned}
\end{equation}
in terms of the intrinsic de~Sitter Killing vectors
\begin{equations}[eq:desitter_killing]
D &\equiv - H^{-1} \partial_t + \vec{x}^i \partial_i &&(\text{dilations}) \eqend{,} \\
K_i &\equiv \left( \vec{x}^2 - H^{-2} \mathe^{- 2 H t} \right) \partial_i - 2 \vec{x}_i D &&(\text{boosts}) \eqend{,} \\
P_i &\equiv \partial_i &&(\text{transl.}) \eqend{,} \\
L_{ij} &\equiv \vec{x}_i \partial_j - \vec{x}_j \partial_i &&(\text{rotations}) \eqend{.}
\end{equations}

\begin{figure}[ht]
\includegraphics[scale=0.55]{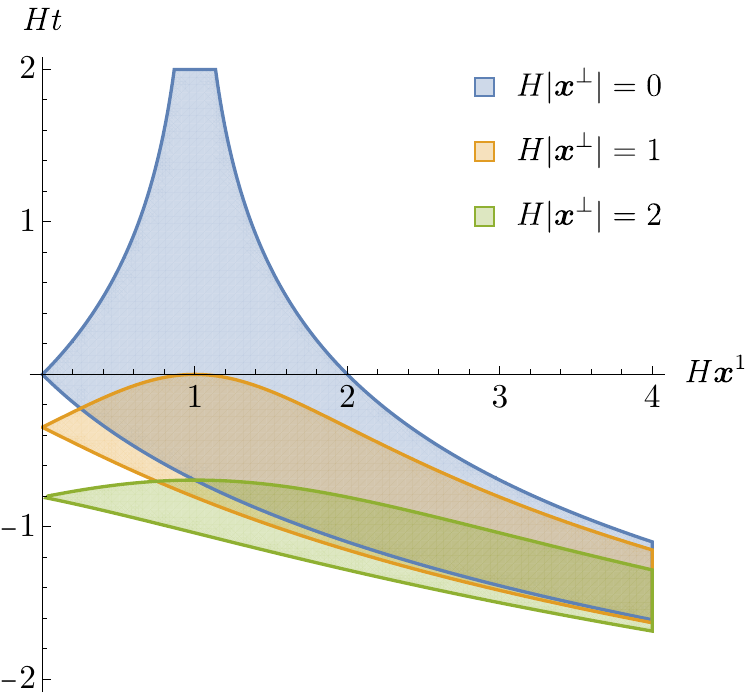}
\caption{Cross sections of the wedge $\mathcal{W}_1$ in the Poincar{\'e} patch of de~Sitter spacetime with coordinates $(t,\vec{x}^1,\vec{x}^\perp)$ for different $\abs{\vec{x}^\perp}$. With growing $\abs{\vec{x}^\perp}$, the cross sections become smaller and shift to earlier times.}
\label{fig:dswedge}
\end{figure}

Following Ref.~\cite{borchersbuchholz1999}, we can then define wedges in $\dS_n$ as the intersection of a wedge in the ambient space with the de~Sitter hyperboloid. In the ambient space, these are the regions $W_j \equiv \{ X \in \mathbb{R}^{n,1} \colon X^j > \abs{X^0} \}$ (for a right wedge in the $j$ direction) or $X^j < - \abs{X^0}$ (for a left wedge), and thus we define the (right) de~Sitter wedges
\begin{equation}
\label{eq:dswedge}
\mathcal{W}_j \equiv \{ X \in \mathbb{R}^{n,1} \colon X^A X_A = H^{-2} \eqend{,} \, X^j > \abs{X^0} \} \eqend{.}
\end{equation}
In the coordinates~\eqref{eq:poincarepatch} of the Poincar{\'e} patch, these are the regions
\begin{equation}
\label{eq:dswedge_poincare}
\mathcal{W}_j = \{ (t,\vec{x})\colon 2 H \vec{x}^j \geq \abs{ 1 - \mathe^{- 2 H t} + H^2 \vec{x}^2 } \}
\end{equation}
which are somewhat difficult to visualize, and we therefore depict them in Fig.~\ref{fig:dswedge}.

\begin{figure}[ht]
\includegraphics[scale=0.55]{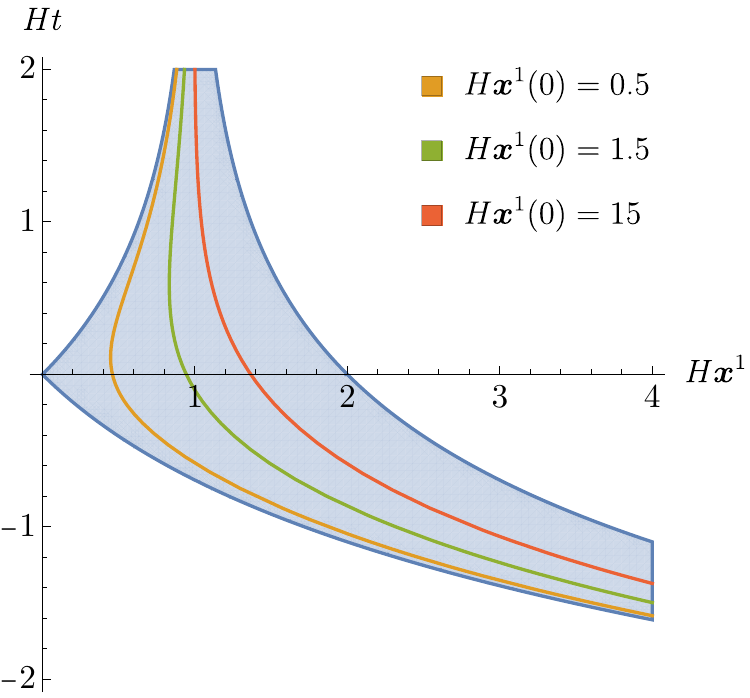}
\caption{Example trajectories with generator $M_{01}$ mapped from the ambient space to the de~Sitter hyperboloid. For simplicity, we chose $\vec{x}^\perp(0) = \vec{0}$.}
\label{fig:dswedgetraj}
\end{figure}

In the ambient space, the boost generators $M_{0j}$ map the wedge $W_j$ into itself. This can be seen easily, taking for example $j = 1$: the ambient wedge $W_1$ is composed of trajectories $X^A(s) = \big( X^0(s), X^1(s), X^\perp \big)$ with
\begin{equations}[eq:wedge_trajectory_X]
X^0(s) &= X^1(0) \sinh(s) \eqend{,} \\
X^1(s) &= X^1(0) \cosh(s) \eqend{,}
\end{equations}
satisfying $X^1(s) > \abs{X^0(s)}$. Their generator is exactly $- M_{01}$, i.e., we have $\partial_s X^A(s) = - M_{01} X^A(s)$, and each point in the wedge lies on one such trajectory. Since the boost generators are tangent to the hyperboloid, they also map the de~Sitter wedges $\mathcal{W}_1$ into themselves, and concretely we obtain the trajectories
\begin{equations}[eq:wedge_trajectory_embedding]
t(s) &= H^{-1} \ln\left[ \frac{1 + H \vec{x}^j(0) \sinh(s)}{\sqrt{ 1 + H^2 \vec{x}^2(0) }} \right] \eqend{,} \\
\vec{x}^j(s) &= \frac{\vec{x}^j(0) \cosh(s)}{1 + H \vec{x}^j(0) \sinh(s)} \eqend{,} \\
\vec{x}^\perp(s) &= \frac{\vec{x}^\perp(0)}{1 + H \vec{x}^j(0) \sinh(s)} \eqend{,}
\end{equations}
with some examples of trajectories depicted in Fig.~\ref{fig:dswedgetraj}.
Using the explicit expression of the Killing vector
\begin{equation}
\label{eq:killing_m0j}
M_{0j} = - \frac{1}{2 H} \left( 1 - \mathe^{- 2 H t} + H^2 \vec{x}^2 \right) \partial_j + H \vec{x}^j \vec{x}^k \partial_k - \vec{x}^j \partial_t \eqend{,}
\end{equation}
it is laborious, but straightforward, to verify that $- M_{0j}$ is their generator, and it is also straightforward to verify that the trajectories stay in the wedge~\eqref{eq:dswedge_poincare} as long as $1 + H \vec{x}^j(0) \sinh(s) \geq 0$. In fact, the range $s \in (s_\text{min},\infty)$ with $s_\text{min} = - \ln\left[ \frac{1 + \sqrt{1 + H^2 [\vec{x}^j(0)]^2}}{H \vec{x}^j(0)} \right]$ already corresponds to the full range $t(s) \in \mathbb{R}$. That a maximum value for $s$ exists in the expanding Poincar{\'e} patch is clear from the condition $X^0 - X^n > 0$, since for negative enough $s$ one obtains $X^0(s) - X^n(s) = X^1(0) \sinh(s) - X^n < 0$, which is a point in the contracting Poincar{\'e} patch. To obtain a better parametrization of the trajectories~\eqref{eq:wedge_trajectory_embedding} in the expanding patch only, we can express $s$ in terms of $t(s)$. This results in
\begin{equations}[eq:wedge_trajectory_intrinsic]
\vec{x}^j(t) &= \mathe^{- H ( t - t_0 )} \sqrt{ \left( \vec{x}^j_0 \right)^2 + \frac{\left( \mathe^{H ( t - t_0 )} - 1 \right)^2}{H^2} } \eqend{,} \\
\vec{x}^\perp(t) &= \mathe^{- H ( t - t_0 )} \vec{x}^\perp_0 \eqend{,}
\end{equations}
where the point $(t_0,\vec{x}^j_0,\vec{x}^\perp_0) \in \mathcal{W}_1$ must be chosen to lie inside the wedge~\eqref{eq:dswedge_poincare}. In this form, the existence of the trajectories for all $t \in \mathbb{R}$ in the expanding Poincar{\'e} patch is now manifest.

Furthermore, left wedges are defined by
\begin{equation}
\label{eq:dswedge_prime}
\mathcal{W}_{-j} \equiv \{ X \in \mathbb{R}^{n,1} \colon X^A X_A = H^{-2} \eqend{,} \, X^j < - \abs{X^0} \} \eqend{,}
\end{equation}
and for them analogous results hold. Of importance for modular theory is the fact that all points in the left wedge $\mathcal{W}_{-j}$ are spacelike related to any point in the right wedge $\mathcal{W}_j$, and that the union $\mathcal{W}_j \cup \mathcal{W}_{-j}$ contains a Cauchy surface for the full de~Sitter space.

\section{Scalar fields and modular theory}
\label{sec:scalarmodular}

We start with a quick review of Tomita--Takesaki modular theory~\cite{tomita1967,takesaki1970}, and refer to Ref.~\cite{takesaki2003} for a thorough mathematical exposition and Ref.~\cite{witten2018} for a more physical one. Modular theory concerns itself with a von~Neumann algebra $\mathcal{A} \subset \mathcal{B}(\mathcal{H})$ of bounded operators acting on a Hilbert space $\mathcal{H}$, and states $\ket{\Psi}$, $\ket{\Phi}$ which are cyclic and separating for $\mathcal{A}$. The state $\ket{\Psi}$ is said to be separating for $\mathcal{A}$ if $a \ket{\Psi} = 0$ for $a \in \mathcal{A}$ implies $a = 0$, and it is cyclic if the set $\{ a \ket{\Psi}\colon a \in \mathcal{A} \} \subset \mathcal{H}$ is a dense subset of the Hilbert space. We also need the commutant algebra $\mathcal{A}'$, which is the set of all operators $a' \in \mathcal{B}(\mathcal{H})$ which commute with all elements in $\mathcal{A}$.

Consider then the map $S_{\Psi\vert\Phi}$ which acts according to $S_{\Psi\vert\Phi} a \ket{\Psi} = a^\dagger \ket{\Phi}$ for $a \in \mathcal{A}$. Tomita proved that this is an unbounded, densely defined and closable operator, and we can take its closure, which we denote by the same symbol. $S_{\Psi\vert\Phi}$ is called the relative Tomita operator, and the adjective ``relative'' is dropped when one considers a single state such that $\Psi = \Phi$. We perform its (unique) polar decomposition $S_{\Psi\vert\Phi} = J_{\Psi\vert\Phi} \Delta^{1/2}_{\Psi\vert\Phi}$ into an antiunitary operator $J_{\Psi\vert\Phi}$ called the relative modular conjugation, and a Hermitean positive-definite operator $\Delta^{1/2}_{\Psi\vert\Phi}$ called the relative modular operator, and again the adjectives ``relative'' are dropped when one considers a single state. The logarithm of the relative modular operator $K_{\Psi\vert\Phi} = \ln \Delta_{\Psi\vert\Phi}$ is known as the relative modular Hamiltonian, since the modular operator induces an automorphism $\alpha$ of $\mathcal{A}$ according to
\begin{equation}
\label{eq:modular_automorphism}
\alpha_s(a) = \Delta^{\mathi s}_{\Psi\vert\Phi} \, a \, \Delta^{- \mathi s}_{\Psi\vert\Phi} = \mathe^{\mathi s K_{\Psi\vert\Phi}} \, a \, \mathe^{- \mathi s K_{\Psi\vert\Phi}} \in \mathcal{A}
\end{equation}
for $a \in \mathcal{A}$, which can be seen as a sort of internal time evolution, called the modular flow. While conjugation by $\Delta^{\mathi s}_{\Psi\vert\Phi}$ thus maps the algebra into itself (and analogously the commutant $\mathcal{A}'$ into itself), conjugation by the relative modular conjugation $J_{\Psi\vert\Phi}$ as well as by $\Delta^{1/2}_{\Psi\vert\Phi}$ exchanges the two algebras. Lastly, we note that the automorphism~\eqref{eq:modular_automorphism} is actually independent of the choice of $\ket{\Psi}$, and that the state $\ket{\Phi}$ is a thermal state with respect to the modular flow, which satisfies the KMS condition.

In our case, we take $\mathcal{A}$ to be the algebra generated by the Weyl operators $W(f) \equiv \mathe^{\mathi \phi(f)}$ of a free massive scalar field, with the test function $f$ restricted to the wedge: $\supp f \subset \mathcal{W}_1$. Since the field $\phi$ is Hermitean, we can further restrict to real $f$, which we will do in the following. Using the Baker--Campbell--Hausdorff (BCH) formula~\cite{achillesbonfiglioli2012} and the commutator~\eqref{eq:commutator_phi_smeared}, one finds that
\begin{splitequation}
W(f) W(h) &= \mathe^{\mathi \phi(f)} \mathe^{\mathi \phi(h)} = \mathe^{\mathi \phi(f) + \mathi \phi(h) + \frac{\mathi}{2} \left( f, \Delta h \right)} \\
&= \mathe^{\frac{\mathi}{2} \left( f, \Delta h \right)} W(f+h) \eqend{,}
\end{splitequation}
such that these really generate an algebra. Moreover, since the commutator function $\Delta$ is antisymmetric one sees immediately that the Weyl operators are invertible with inverse $[ W(f) ]^{-1} = W(-f)$. The Hilbert space $\mathcal{H}$ is nothing else but the Fock space of the scalar field, and the de~Sitter-invariant Bunch--Davies vacuum state~\cite{chernikovtagirov1968,schomblondspinde1976,bunchdavies1978} whose construction is reviewed in Appendix~\ref{sec:app_quantization} is cyclic and separating~\cite{borchersbuchholz1999}. Since the Weyl operators are invertible, also the coherent states $\ket{f} \equiv W(f) \ket{\Omega}$ (without restriction on the support of $f$) are cyclic and separating, and one can thus define relative modular operators for them.

In fact, let us take $\ket{\Psi} = \ket{f}$ and $\ket{\Phi} = \ket{h}$ for two test functions $f$ and $h$. We use the representation of the smeared field $\phi(f)$ in terms of initial data~\eqref{eq:phi_smeared_symplectic}
\begin{equation}
\phi(f) = - \mathi \symp{ \phi, \Delta f } \eqend{,}
\end{equation}
where $\symp{ \cdot, \cdot }$ is the symplectic product~\eqref{eq:symplectic_product} supported on a Cauchy surface $\Sigma$, and choose $\Sigma$ contained in the union $\mathcal{W}_1 \cup \mathcal{W}_{-1}$. This Cauchy surface thus splits into two parts contained in the respective wedges, $\Sigma = \Sigma_1 \cup \Sigma_{-1}$ with $\Sigma_1 \subset \mathcal{W}_1$ and $\Sigma_{-1} \subset \mathcal{W}_{-1}$. Accordingly, we split the smeared commutator on the Cauchy surface $\Delta f \big\rvert_\Sigma$ into two parts $(\Delta f)_+$ and $(\Delta f)_-$ whose support lies in the respective half of the Cauchy surface, i.e., such that $(\Delta f)_+ \big\rvert_{\Sigma_{-1}} = 0$ and $(\Delta f)_- \big\rvert_{\Sigma_1} = 0$. It follows that
\begin{splitequation}
\label{eq:weyl_decomposition}
W(f) &= \mathe^{\symp{ \phi, \Delta f }} \\
&= \mathe^{\symp{ \phi, (\Delta f)_+ }} \mathe^{\symp{ \phi, (\Delta f)_- }} \equiv U_+(f) U_-(f) \eqend{,}
\end{splitequation}
where we used the BCH formula, the commutator~\eqref{eq:commutator_symplectic}, and the fact that
\begin{equation}
\symp{ (\Delta f)_+, (\Delta f)_- } = 0 \eqend{,}
\end{equation}
since the supports of $(\Delta f)_+$ and $(\Delta f)_-$ are disjoint by construction. It follows that also $[ U_+(f), U_-(f) ] = 0$, and using that $W(f)^\dagger = W(-f)$ also $U_\pm(f)^\dagger = U_\pm(-f)$. By construction, $U_+(f) \in \mathcal{A}$, while the other unitary $U_-(f)$ is an element of the algebra of fields restricted to the left wedge $\mathcal{W}_{-1}$.\footnote{This is the same decomposition that is used in the more mathematical literature~\cite{figlioliniguido1989,ciollilongoruzzi2020,bostelmanncadamurominz2023}, and which was derived in Ref.~\cite{casinigrillopontello2019} in more detail.} In Ref.~\cite{borchersbuchholz1999} it is shown that wedge duality holds, hence this algebra is the commutant of the algebra of fields restricted to the right wedge and $U_-(f) \in \mathcal{A}'$.

We are thus in the situation $\ket{\Psi} = U_+(f) U_-(f) \ket{\Omega}$, $\ket{\Phi} = U_+(h) U_-(h) \ket{\Omega}$ with $U_+ \in \mathcal{A}$ and $U_- \in \mathcal{A}'$. The relative Tomita operator $S_{\Psi\vert\Phi}$ acts according to
\begin{splitequation}
S_{\Psi\vert\Phi} a U_+(f) U_-(f) \ket{\Omega} &= a^\dagger U_+(h) U_-(h) \ket{\Omega} \\
&= U_-(h) \big[ U_+(-h) a \big]^\dagger \ket{\Omega}
\end{splitequation}
for $a \in \mathcal{A}$, where we used that $U_-$ commutes with both $a$ and $U_+$ and that $U_+(h)^\dagger = U_+(-h)$. Multiplying with $U_-(h)^\dagger = U_-(-h)$ and then with $U_+(f)^\dagger = U_+(-f)$, we obtain further
\begin{splitequation}
&U_+(-f) U_-(-h) S_{\Psi\vert\Phi} U_-(f) a U_+(f) \ket{\Omega} \\
&\quad= \big[ U_+(-h) a U_+(f) \big]^\dagger \ket{\Omega} \eqend{,}
\end{splitequation}
where we also used that $U_-(f)$ commutes with both $U_+(f)$ and $a$. By definition of the Tomita operator for the state $\ket{\Omega}$, we have $\big[ U_+(-h) a U_+(f) \big]^\dagger \ket{\Omega} = S_0 U_+(-h) a U_+(f) \ket{\Omega}$, and hence
\begin{splitequation}
&U_+(-f) U_-(-h) S_{\Psi\vert\Phi} U_-(f) a U_+(f) \ket{\Omega} \\
&\quad= S_0 U_+(-h) a U_+(f) \ket{\Omega} \eqend{.}
\end{splitequation}
Since $\ket{\Omega}$ is separating and this equation needs to hold for all $a \in \mathcal{A}$, we can read off that $U_+(-f) U_-(-h) S_{\Psi\vert\Phi} U_-(f) = S_\Omega U_+(-h)$, or
\begin{equation}
S_{\Psi\vert\Phi} = U_-(h) U_+(f) S_0 U_+(-h) U_-(-f) \eqend{.}
\end{equation}
Using the Hermitean properties of the modular conjugation $J$ and the modular operator $\Delta$, it also follows that
\begin{splitequation}
\label{eq:relative_delta_unitary}
\Delta_{\Psi\vert\Phi} &= S_{\Psi\vert\Phi}^\dagger S_{\Psi\vert\Phi}^{\phantom{\dagger}} = U_-(f) U_+(h) S_\Omega^\dagger S_\Omega^{\phantom{\dagger}} U_+(-h) U_-(-f) \\
&= U_-(f) U_+(h) \Delta_\Omega U_+(-h) U_-(-f) \eqend{.} \raisetag{1.3em}
\end{splitequation}

It thus only remains to determine the modular operator $\Delta_\Omega$ for the de~Sitter vacuum state $\ket{\Omega}$, or alternatively the modular Hamiltonian $K_\Omega = \ln \Delta_\Omega$. Instead of using its definition, however, we show that the flow of the two-point function along the trajectories~\eqref{eq:wedge_trajectory_embedding} satisfies the KMS condition. Since the modular flow is the unique one for which this is true, its generator is guaranteed to be the modular Hamiltonian $K_\Omega$, up to a rescaling to fix the (modular) temperature at $\beta = 1$. Consider thus the flowed two-point function
\begin{equation}
F(s,s') \equiv \bra{\Omega} \phi(x(s)) \phi(x'(s')) \ket{\Omega} \eqend{,}
\end{equation}
where the trajectory $x(s)$ is given in Eq.~\eqref{eq:wedge_trajectory_embedding}. The explicit form of the two-point function is given in Eq.~\eqref{eq:wightman_2pf}, and depends on the causal relation between $x$ and $x'$ and the de~Sitter-invariant distance $Z(x,x')$~\eqref{eq:z_def}. For our purposes, it is easier to use the expression~\eqref{eq:z_embedding} of $Z$ in terms of the embedding coordinates $X^A$, since those have a simple expression~\eqref{eq:wedge_trajectory_X} along the trajectories. We obtain
\begin{splitequation}
Z(x(s),x'(s')) &= H^2 X^1(x) X^1(x') \cosh(s-s') \\
&\quad+ H^2 X^\perp(x) \cdot X^\perp(x') \eqend{,}
\end{splitequation}
where we wrote $x = x(0)$ and $x' = x'(0)$ to shorten the expression. If the points $x(s)$ and $x'(s)$ are spacelike separated (which entails $Z(x,x') < 1$) such that the $\mathi \epsilon$ prescription in the two-point function~\eqref{eq:wightman_2pf} is irrelevant, we thus obtain $F(s,s') = F(s-s') = F(s'-s)$. On the other hand, if they are timelike separated (such that $Z(x,x') > 1$) this prescription determines on which side of the branch cut one has to evaluate the hypergeometric function, resulting in the expression~\eqref{eq:imu_integral_timelike}. We can combine both by considering instead $Z_\epsilon$~\eqref{eq:zepsilon_def}, for which we compute
\begin{splitequation}
&Z_\epsilon(x(s),x'(s')) = Z(x(s),x'(s')) - H^2 \epsilon^2 \\
&\quad- \mathi \epsilon H^2 \big[ X^1(x) \sinh(s) - X^1(x') \sinh(s') \big] \eqend{.}
\end{splitequation}

The KMS condition~\cite{kubo1957,martinschwinger1959,haaghugenholtzwinnink1967} states that the the function $F$ can be analytically continued in some strip parallel to the real axis, and when the differences of the imaginary parts of their arguments are equal to the inverse temperature $\beta$ one obtains the complex conjugate function. In our case, we see that
\begin{equation}
Z_\epsilon(x(s-\mathi \pi),x'(s'+\mathi \pi)) = Z_\epsilon^*(x(s),x'(s')) \eqend{,}
\end{equation}
and since the hypergeometric function (off the branch cut) is an analytic function of its argument, we obtain $F(s-\mathi\pi,s'+\mathi\pi) = F^*(s,s')$\footnote{The other possible analytic continuation $s \to s + \mathi \pi$, $s' \to s' - \mathi \pi$ would cross the branch cut and is thus not permissible.}. It follows that with respect to the evolution whose generator is $- M_{0j}$, the de~Sitter Bunch--Davies vacuum state has temperature $\beta = 2 \pi$. Rescaling to obtain $\beta = 1$ and using that each Killing vector corresponds to a self-adjoint operator on the Fock space of the free scalar field which we denote by calligraphic letters~\eqref{eq:killing_op_commutator}, it follows that
\begin{equation}
\label{eq:modular_hamiltonian}
\ln \Delta_\Omega = K_\Omega = - 2 \pi \mathcal{M}_{0j} = \frac{\pi}{H} \left( H^2 \mathcal{K}_j + \mathcal{P}_j \right) \eqend{,}
\end{equation}
which is the result previously derived in Ref.~\cite{borchersbuchholz1999}.

\section{Relative entropy}
\label{sec:entropy}

With the modular Hamiltonian~\eqref{eq:modular_hamiltonian} at our disposal, we can now use the Araki--Uhlmann formula~\eqref{eq:araki_uhlmann} to compute the relative entropy between two coherent excitations $\ket{f} \equiv W(f) \ket{\Omega}$ and $\ket{h}$. We decompose the Weyl operators $W(f)$ and $W(h)$ according to Eq.~\eqref{eq:weyl_decomposition}, and use the formula~\eqref{eq:relative_delta_unitary} for the relative modular operator. Because conjugation by unitary operators extends to functions of an operator, we also have
\begin{equation}
\ln \Delta_{f\vert h} = U_-(f) U_+(h) \ln \Delta_\Omega \, U_+(-h) U_-(-f) \eqend{,}
\end{equation}
and hence
\begin{equation}
\label{eq:relative_entropy_weyl}
\mathcal{S}(f\Vert h) = - \bra{\Omega} U_+(-f) U_+(h) \ln \Delta_\Omega \, U_+(-h) U_+(f) \ket{\Omega} \eqend{,}
\end{equation}
which coincides with Eq.~\eqref{eq:araki_uhlmann_coherent} given in the introduction since $U_+(f) \in \mathcal{A}$.

Using the BCH formula and the commutator~\eqref{eq:commutator_symplectic}, we first compute
\begin{splitequation}
U_+(-h) U_+(f) &= \mathe^{- \symp{ \phi, (\Delta h)_+ }} \mathe^{\symp{ \phi, (\Delta f)_+ }} \\
&= \mathe^{\symp{ \phi, (\Delta (f-h))_+ }} \mathe^{\frac{1}{2} \symp{ (\Delta h)_+, (\Delta f)_+ }} \\
&= U_+(f-h) \, \mathe^{\frac{1}{2} \symp{ (\Delta h)_+, (\Delta f)_+ }} \eqend{,}
\end{splitequation}
such that Eq.~\eqref{eq:relative_entropy_weyl} reduces to
\begin{equation}
\label{eq:relative_entropy_weyl_2}
\mathcal{S}(f\Vert h) = - \bra{\Omega} U_+(h-f) \ln \Delta_\Omega \, U_+(f-h) \ket{\Omega} \eqend{,}
\end{equation}
taking into account the antisymmetry of the symplectic product $\sigma$ when we exchange its arguments. We see that only the difference of test functions enters, such that $\mathcal{S}(f\Vert h) = \mathcal{S}(f-h\Vert \Omega)$, the relative entropy between the vacuum $\ket{\Omega}$ and the coherent excitation $W(f-h) \ket{\Omega}$. We thus may and will set $h = 0$ in the following. The same simplification happens in Minkowski space~\cite{casinigrillopontello2019}, and in fact is a general property of the relative entropy for coherent excitations~\cite{ciollilongoruzzi2020}.

To compute this expectation value, we employ the identity
\begin{equation}
\label{eq:identity_adjoint_action}
\mathe^{- \mathi v B} A \mathe^{\mathi v B} = A + \mathi \int_0^v \mathe^{- \mathi u B} [A,B] \mathe^{\mathi u B} \total u \eqend{,}
\end{equation}
which is easily proven by noting that it holds for $v = 0$, and that both sides have the same derivative with respect to $v$. From the explicit expression for the modular Hamiltonian~\eqref{eq:modular_hamiltonian}, we know that
\begin{equation}
\mathi \big[ \ln \Delta_\Omega, \phi(x) \big] = - 2 \pi M_{0j} \phi(x) \eqend{,}
\end{equation}
or after smearing with a test function and using the symmetry properties~\eqref{eq:killing_inner_product}
\begin{splitequation}
\mathi \big[ \ln \Delta_\Omega, \phi(f) \big] &= - 2 \pi \left( f, M_{0j} \phi \right) \\
&= 2 \pi \left( M_{0j} f, \phi \right) = 2 \pi \phi(M_{0j} f) \eqend{,}
\end{splitequation}
which is linear in $\phi$. Therefore, its commutator with $\phi$ is proportional to the identity operator, such that using~\eqref{eq:identity_adjoint_action} twice we obtain
\begin{splitequation}
\mathe^{- \mathi v \phi(f)} \ln \Delta_\Omega \, \mathe^{\mathi v \phi(f)} &= \ln \Delta_\Omega + \mathi v \big[ \ln \Delta_\Omega, \phi(f) \big] \\
&\quad- \frac{v^2}{2} \big[ \big[ \ln \Delta_\Omega, \phi(f) \big], \phi(f) \big] \eqend{.}
\end{splitequation}
Since the vacuum is invariant under the de~Sitter symmetries (i.e., the generators $\mathcal{K}_j$ and $\mathcal{P}_j$ are normal-ordered operators annihilating $\ket{\Omega}$), we have $\bra{\Omega} \ln \Delta_\Omega \ket{\Omega} = 0$, and also $\bra{\Omega} \phi(x) \ket{\Omega} = 0$ from the expansion~\eqref{eq:phi_modes}. Setting $v = 1$, it follows that
\begin{splitequation}
\label{eq:expectation_lnomega_symplectic}
\bra{\Omega} \mathe^{- \mathi \phi(f)} \ln \Delta_\Omega \, \mathe^{\mathi \phi(f)} \ket{\Omega} &= - \frac{1}{2} \big[ \big[ \ln \Delta_\Omega, \phi(f) \big], \phi(f) \big] \\
&= - \mathi \pi \symp{ \Delta M_{0j} f, \Delta f } \eqend{,} \raisetag{1.3em}
\end{splitequation}
where we also used the expression~\eqref{eq:commutator_phi_symplectic} for the commutator in terms of the symplectic product.

Using the initial-value formulation~\eqref{eq:phi_smeared_symplectic} of the smeared field to write $\phi(f) = - \mathi \symp{ \phi, \Delta f}$, Eq.~\eqref{eq:expectation_lnomega_symplectic} is almost in a form that can be used for the relative entropy~\eqref{eq:relative_entropy_weyl_2}. The only missing part is to express $\Delta M_{0j} f$ as a function of $\Delta f$. For this, we compute using the symmetry properties of $M_{0j}$~\eqref{eq:killing_inner_product} that
\begin{splitequation}
(\Delta M_{0j} f)(x) &= \left( \Delta(x,\cdot), M_{0j} f \right) \\
&= - \left( M_{0j} \Delta(x,\cdot), f \right) \eqend{,}
\end{splitequation}
where in the last expression $M_{0j}$ acts on the second argument of $\Delta$. Using the results~\eqref{eq:delta_covariance_p} and~\eqref{eq:delta_covariance_k}, it then follows that $\left( M_{0j} \Delta(x,\cdot), f \right) = - M_{0j} \left( \Delta(x,\cdot), f \right)$ with $M_{0j}$ acting now on the external point $x$, and thus
\begin{equation}
(\Delta M_{0j} f)(x) = M_{0j} \left( \Delta(x,\cdot), f \right) = (M_{0j} \Delta f)(x) \eqend{.}
\end{equation}
Eq.~\eqref{eq:expectation_lnomega_symplectic} therefore simplifies to
\begin{equation}
\bra{\Omega} \mathe^{- \symp{ \phi, \Delta f }} \ln \Delta_\Omega \, \mathe^{\symp{ \phi, \Delta f }} \ket{\Omega} = - \mathi \pi \symp{ M_{0j} \Delta f, \Delta f } \eqend{,}
\end{equation}
and replacing $\Delta f$ by $(\Delta f)_+$ we obtain our first result for the relative entropy~\eqref{eq:relative_entropy_weyl_2}:
\begin{equation}
\label{eq:relative_entropy_weyl_3}
\mathcal{S}(f\Vert \Omega) = \mathi \pi \symp{ M_{0j} (\Delta f)_+, (\Delta f)_+ } \eqend{.}
\end{equation}
From this expression and the previously derived $\mathcal{S}(f\Vert h) = \mathcal{S}(f-h\Vert \Omega)$, we already see that the relative entropy is symmetric: $\mathcal{S}(f\Vert \Omega) = \mathcal{S}(\Omega \Vert f)$. In particular, the relative entropy between the vacuum and a coherent excitation is the same as between a coherent excitation and the vacuum.

A more explicit expression for the relative entropy can be obtained by evaluating the sympletic product in~\eqref{eq:relative_entropy_weyl_3} explicitly. On the Cauchy surface $t = 0$ that is adapted to the Poincar{\'e} patch, and with the explicit expression~\eqref{eq:killing_m0j} for $M_{0j}$, we obtain
\begin{widetext}
\begin{splitequation}
\label{eq:relative_entropy_cauchy_1}
\mathcal{S}(f\Vert \Omega) &= \pi \int \Big[ \vec{x}^j \partial_t (\Delta f)_+(x) \partial_t (\Delta f)_+(x) - \vec{x}^j (\Delta f)_+(x) \partial_t^2 (\Delta f)_+(x) + \frac{H}{2} \vec{x}^2 \partial_j (\Delta f)_+(x) \partial_t (\Delta f)_+(x) \\
&\qquad\quad- \frac{H}{2} \vec{x}^2 (\Delta f)_+(x) \partial_j \partial_t (\Delta f)_+(x) - H \vec{x}^j \vec{x}^k \partial_k (\Delta f)_+(x) \partial_t (\Delta f)_+(x) \\
&\qquad\quad+ H \vec{x}^j \vec{x}^k (\Delta f)_+(x) \partial_k \partial_t (\Delta f)_+(x) - (\Delta f)_+(x) \partial_j (\Delta f)_+(x) \Big]_{t = 0} \total^{n-1} \vec{x} \eqend{.}
\end{splitequation}
\end{widetext}
To simplify this expression, we first have to consider the precise form of the splitting of the smeared commutator into the parts $(\Delta f)_+$ and $(\Delta f)_-$. Setting $t = 0$ in the wedge $\mathcal{W}_j$~\eqref{eq:dswedge_poincare}, we see that the intersection of the Cauchy surface with $\mathcal{W}_j$ is the region
\begin{splitequation}
\label{eq:cauchy_wedge_intersection}
\mathcal{R}_j &= \left\{ (0,\vec{x})\colon 2 \vec{x}^j \geq H \vec{x}^2 \right\} \\
&= \left\{ (0,\vec{x})\colon H \abs{ \vec{x}^\perp } \leq \sqrt{ 1 - \left( 1 - H \vec{x}^j \right)^2 } \right\} \eqend{.}
\end{splitequation}
Defining the indicator function
\begin{equation}
\label{eq:indicator_chi}
\chi_j(x) \equiv \left. \begin{cases} 1 & x \in \mathcal{R}_j \\ 0 & x \notin \mathcal{R}_j \end{cases} \right\} = \Theta\big( 2 \vec{x}^j - H \vec{x}^2 \big) \eqend{,}
\end{equation}
we therefore set $( \Delta f )_+(x) \equiv \chi_j(x) ( \Delta f )(x)$. Using that $\Delta$ fulfills the Klein--Gordon equation~\eqref{eq:klein_gordon} and integrating spatial derivatives by parts, Eq.~\eqref{eq:relative_entropy_cauchy_1} can then be simplified to
\begin{splitequation}
\label{eq:relative_entropy_cauchy_2}
\mathcal{S}(f\Vert \Omega) &= 2 \pi \int \chi_j(x) \vec{x}^j T_{00}(\Delta f) \total^{n-1} \vec{x} \\
&\quad+ \pi H \int \chi_j(x) \left( \vec{x}^2 \delta_j^k - 2 \vec{x}^j \vec{x}^k \right) \\
&\qquad\quad\times \partial_k (\Delta f)(x) \partial_t (\Delta f)(x) \total^{n-1} \vec{x} \\
&\quad+ \mathcal{S}_\text{bdy}(f\Vert \Omega) \eqend{,}
\end{splitequation}
where
\begin{equation}
\label{eq:energy_density}
T_{00}(h) \equiv \frac{1}{2} \left( \partial_t h \right)^2 + \frac{1}{2} \partial^k h \partial_k h + \frac{1}{2} m^2 h^2
\end{equation}
is the classical energy density of a free scalar field on the Cauchy surface $t = 0$. The last term $\mathcal{S}_\text{bdy}(f\Vert \Omega)$ are contributions located at the boundary of the region~\eqref{eq:cauchy_wedge_intersection} coming from integration by parts. Their treatment is somewhat delicate, since naively terms appear which are not uniquely defined. In fact, with a smoothed version of the indicator function~\eqref{eq:indicator_chi} a direct computation gives
\begin{splitequation}
\label{eq:entropy_boundary}
\mathcal{S}_\text{bdy}(f\Vert \Omega) &= - \pi \int \bigg[ \frac{H}{2} \left( 2 \vec{x}^j \vec{x}^k - \vec{x}^2 \delta^k_j \right) \partial_t \big[ (\Delta f)(x) \big]^2 \\
&\qquad+ \delta^k_j \big[ (\Delta f)(x) \big]^2 \bigg]_{t = 0} \! \chi_j(x) \partial_k \chi_j(x) \total^{n-1} \vec{x} \eqend{.}
\end{splitequation}

In the flat-space limit $H \to 0$, we have $\chi_j(x) \to \Theta(\vec{x}^j)$ such that
\begin{equation}
\label{eq:entropy_boundary_minkowski}
\lim_{H \to 0} \mathcal{S}_\text{bdy}(f\Vert \Omega) = - \pi \Theta(0) \int \big[ (\Delta f)(x) \big]^2_{t,\vec{x}^j = 0} \total^{n-2} \vec{x}^\perp \eqend{,}
\end{equation}
and the boundary contribution to the relative entropy depends on the value of the regularized $\Theta(0)$. To determine the correct value, one has to delve into the mathematical details of convergence in an appropriate function space, which was done in Ref.~\cite{casinigrillopontello2019} for the Minkowski case. There it was shown that one needs to exhaust the wedge region from within, i.e., take an approximation for which $\Theta(0) = 0$ such that the boundary terms~\eqref{eq:entropy_boundary_minkowski} vanish. Generalizing their arguments to the de~Sitter case, we have to take an approximation for which $\chi_j(x) \partial_k \chi_j(x) \to 0$ such that the boundary terms~\eqref{eq:entropy_boundary} vanish in general and $\mathcal{S}_\text{bdy}(f\Vert \Omega) = 0$.

While the classical energy density~\eqref{eq:energy_density} is clearly positive and $\vec{x}^j \geq 0$ in the region $\mathcal{R}_j$~\eqref{eq:cauchy_wedge_intersection} such that the first contribution to the relative entropy~\eqref{eq:relative_entropy_cauchy_2} is positive, already the second term does not have a definite sign. Since general considerations~\cite{witten2018,ciollilongoruzzi2020} enforce the positivity of the relative entropy, we have to conclude that the Cauchy surface $t = 0$ is not suitable to obtain a manifestly positive expression for the relative entropy. Nevertheless, we have already the correct flat-space limit~\cite{casinigrillopontello2019}
\begin{equation}
\label{eq:relative_entropy_cauchy_flat}
\lim_{H \to 0} \mathcal{S}(f\Vert \Omega) = 2 \pi \int_{\vec{x}^j \geq 0} \vec{x}^j T_{00}(\Delta f) \Big\rvert_{t = 0} \total^{n-1} \vec{x} \eqend{.}
\end{equation}
However, since the symplectic product~\eqref{eq:symplectic_product_poincare} is independent of the choice of Cauchy surface, we can evaluate it on any other such surface. In particular, we can choose the surface $X^0 = 0$, which is adapted to the Wedge $\mathcal{W}_j$. In the coordinates~\eqref{eq:poincarepatch} of the Poincar{\'e} patch, this is the surface~\eqref{eq:cauchy_surface_tilted}
\begin{equation}
\label{eq:cauchy_surface_tilted_main}
\Sigma = \left\{ (t,\vec{x})\colon 2 H t + \ln\left( 1 + H^2 \vec{x}^2 \right) = 0 \right\} \eqend{,}
\end{equation}
which has the future-pointing unit normal vector~\eqref{eq:normal_vector_tilted_surface}
\begin{equation}
\label{eq:normal_vector_tilted_surface_main}
n^\mu \big\rvert_\Sigma = \sqrt{1 + H^2 \vec{x}^2} \left( 1, - H \vec{x}^i \right)^\mu
\end{equation}
and the induced metric~\eqref{eq:induced_metric_tilted}
\begin{equation}
\gamma_{ij} = \frac{\delta_{ij}}{1 + H^2 \vec{x}^2} - \frac{H^2 \vec{x}_i \vec{x}_j}{( 1 + H^2 \vec{x}^2 )^2} \eqend{.}
\end{equation}

The sympletic product on this Cauchy surface is given in Eq.~\eqref{eq:symplectic_product_tilted}, such that we can evaluate the relative entropy~\eqref{eq:relative_entropy_weyl_3} using also the identities~\eqref{eq:m0j_commutator_tilted} and~\eqref{eq:boundary_tilted}. This results in
\begin{widetext}
\begin{splitequation}
\label{eq:relative_entropy_cauchy_tilted}
\mathcal{S}(f\Vert \Omega) &= \pi \int \frac{\vec{x}^j}{\sqrt{ 1 + H^2 \vec{x}^2 }} \bigg[ \partial_n (\Delta f)_+(x) \partial_n (\Delta f)_+(x) + \big[ ( 1 + H^2 \vec{x}^2 ) \delta^{kl} - H^2 \vec{x}^k \vec{x}^l \big] \hat\partial_k (\Delta f)_+(x) \hat\partial_l (\Delta f)_+(x) \\
&\hspace{9em}+ m^2 \big[ (\Delta f)_+(x) \big]^2 + (\Delta f)_+(x) \left( \nabla^2 - m^2 \right) (\Delta f)_+(x) \bigg]_\Sigma \sqrt{\gamma} \total^{n-1} \vec{x}
\end{splitequation}
\end{widetext}
with $\partial_n \equiv n^\mu \partial_\mu$ and $\hat\partial_\mu \equiv \partial_\mu + n_\mu \partial_n$, which still can be simplified. For this, we again need to be precise about the splitting of the smeared commutator. Since the Cauchy surface $X^0 = 0$ is adapted to the wedge $\mathcal{W}_j$, this is easier than before, and computing the intersection of $\Sigma$~\eqref{eq:cauchy_surface_tilted_main} with the wedge $\mathcal{W}_j$~\eqref{eq:dswedge_poincare}, we see that we can simply obtain $\vec{x}^j \geq 0$ and hence $\chi_j(x) = \Theta(\vec{x}^j)$. Using the explicit expressions~\eqref{eq:derivatives_tilted_sigma} for $\partial_n$ and $\hat\partial_k$ on the Cauchy surface $\Sigma$, we compute
\begin{equations}
\partial_n (\Delta f)_+(x) &= \chi_j(x) \partial_n (\Delta f)(x) \eqend{,} \\
\hat\partial_k (\Delta f)_+(x) &= \chi_j(x) \hat\partial_k (\Delta f)(x) + \delta_k^j (\Delta f)(x) \delta(\vec{x}^j) \eqend{,}
\end{equations}
and using also the Klein--Gordon equation~\eqref{eq:klein_gordon_tilted} we compute furthermore
\begin{splitequation}
&\left( \nabla^2 - m^2 \right) (\Delta f)_+(x) = \chi_j(x) \left( \nabla^2 - m^2 \right) (\Delta f)(x) \\
&\quad+ \left( 1 + H^2 \vec{x}^2 \right) \left[ \delta'(\vec{x}^j) (\Delta f)(x) + 2 \delta(\vec{x}^j) \partial_j (\Delta f)(x) \right] \eqend{,}
\end{splitequation}
where all expressions are understood to be restricted to the Cauchy surface $\Sigma$. The first term on the right-hand side vanishes since the commutator function $\Delta$ satisfies the Klein--Gordon equation, while the other terms result in boundary contributions. Inserting these results into Eq.~\eqref{eq:relative_entropy_cauchy_tilted}, we obtain
\begin{equation}
\label{eq:relative_entropy_cauchy_tilted_2}
\mathcal{S}(f\Vert \Omega) = 2 \pi \int \chi_j(x) \mathcal{Q}_j(\Delta f) \big\rvert_\Sigma \sqrt{\gamma} \total^{n-1} \vec{x} + \mathcal{S}_\text{bdy}(f\Vert \Omega) \eqend{,}
\end{equation}
where we defined
\begin{splitequation}
\label{eq:q_def}
\mathcal{Q}_j(h) \big\rvert_\Sigma &\equiv \frac{\vec{x}^j}{2 \sqrt{ 1 + H^2 \vec{x}^2 }} \Big[ \partial_n h \partial_n h + m^2 h^2 \\
&\quad+ \big[ ( 1 + H^2 \vec{x}^2 ) \delta^{kl} - H^2 \vec{x}^k \vec{x}^l \big] \hat\partial_k h \hat\partial_l h \Big]_\Sigma \eqend{,} \raisetag{4.2em}
\end{splitequation}
and the boundary terms are formally given by
\begin{splitequation}
\mathcal{S}_\text{bdy}(f\Vert \Omega) &= \pi \int \vec{x}^j \sqrt{ 1 + H^2 \vec{x}^2 } \bigg[ \big[ (\Delta f)(x) \delta(\vec{x}^j) \big]^2 \\
&\qquad+ 2 \chi_j(x) \delta(\vec{x}^j) \partial_j \big[ (\Delta f)(x) \big]^2 \\
&\qquad+ \chi_j(x) \delta'(\vec{x}^j) \big[ (\Delta f)(x) \big]^2 \bigg]_\Sigma \sqrt{\gamma} \total^{n-1} \vec{x} \eqend{.}
\end{splitequation}
We see that they are again undefined since $\chi_j(x) \delta'(\vec{x}^j)$ can take arbitrary values depending on the regularization. Moreover, also divergent terms appear since $[ \delta(\vec{x}^j) ]^2 = \infty$, but there is extra $\vec{x}^j$ in front of everything which lowers the degree of divergence. To obtain a proper result, we have again to take a smoothed-out version of the indicator function $\chi_j$, which then results in
\begin{splitequation}
\label{eq:entropy_boundary_tilted}
\mathcal{S}_\text{bdy}(f\Vert \Omega) &= - \pi \int \sqrt{ 1 + H^2 \vec{x}^2 } \big[ (\Delta f)(x) \big]^2_\Sigma \\
&\qquad\times \chi_j(x) \partial_j \chi_j(x) \sqrt{\gamma} \total^{n-1} \vec{x}
\end{splitequation}
up to terms that vanish for any smoothing. Without smoothing, we would obtain $\chi_j(x) \partial_j \chi_j(x) = \Theta(0) \delta(\vec{x}^j)$, which clearly has the same flat-space limit~\eqref{eq:entropy_boundary_minkowski} as the result for the other Cauchy surface at $t = 0$. As in the flat-space case~\cite{casinigrillopontello2019}, the proper smoothing is such that one exhausts the wedge from within, taking an approximation for which $\chi_j(x) \partial_j \chi_j(x) \to 0$. The boundary terms~\eqref{eq:entropy_boundary_tilted} then vanish and $\mathcal{S}_\text{bdy}(f\Vert \Omega) = 0$ as before, such that only the bulk terms involving $\mathcal{Q}_j$~\eqref{eq:q_def} contribute to the relative entropy~\eqref{eq:relative_entropy_cauchy_tilted_2}.

We have thus obtained
\begin{equation}
\label{eq:relative_entropy_cauchy_tilted_3}
\mathcal{S}(f\Vert \Omega) = 2 \pi \int \Theta(\vec{x}^j) \mathcal{Q}_j(\Delta f) \big\rvert_\Sigma \sqrt{\gamma} \total^{n-1} \vec{x}
\end{equation}
with $\mathcal{Q}_j(h)$ given by Eq.~\eqref{eq:q_def}. Since for positive $\vec{x}^j$ we have $\mathcal{Q}_j(h) \geq 0$\footnote{The matrix $\big[ ( 1 + H^2 \vec{x}^2 ) \delta^{kl} - H^2 \vec{x}^k \vec{x}^l \big]$ is positive definite, since its eigensystem is given by a single eigenvector $\vec{x}_k$ with eigenvalue $1$ and $(n-2)$ transverse eigenvectors $( \delta_{kl} \vec{x}^2 - \vec{x}_k \vec{x}_l ) v^l$ with eigenvalue $( 1 + H^2 \vec{x}^2 )$ for constant $v^l$, so all eigenvalues are strictly positive.} for any test function $h$ with equality only if $h = 0$, the positivity of the relative entropy is now manifest: $\mathcal{S}(f\Vert \Omega) \geq 0$ and $\mathcal{S}(f\Vert \Omega) = 0$ only if $f = 0$. It is also easy to see that it is jointly convex, i.e., that
\begin{equation}
\label{eq:relative_entropy_convex}
\lambda \mathcal{S}(f \Vert \Omega) + (1-\lambda) \mathcal{S}(h \Vert \Omega) \geq \mathcal{S}(\lambda f + (1-\lambda) h \Vert \Omega)
\end{equation}
for all test functions $f,h$ and $\lambda \in [0,1]$. However, this is immediate since the relative entropy $\mathcal{S}(f \Vert \Omega)$ is quadratic in the test function $f$, such that
\begin{splitequation}
&\mathcal{S}(\lambda f + (1-\lambda) h \Vert \Omega) = \lambda^2 \mathcal{S}(f \Vert \Omega) + (1-\lambda)^2 \mathcal{S}(h \Vert \Omega) \\
&\qquad+ \lambda (1-\lambda) \big[ \mathcal{S}(f \Vert \Omega) + \mathcal{S}(h \Vert \Omega) - \mathcal{S}(f-h \Vert \Omega) \big] \\
&\quad= \lambda \mathcal{S}(f \Vert \Omega) + (1-\lambda) \mathcal{S}(h \Vert \Omega) - \lambda (1-\lambda) \mathcal{S}(f-h \Vert \Omega) \eqend{.}
\end{splitequation}
Positivity of the relative entropy shows that the last term is negative, $- \lambda (1-\lambda) \mathcal{S}(f-h \Vert \Omega) \leq 0$, and the joint convexity~\eqref{eq:relative_entropy_convex} follows, with a strict inequality if $f \neq h$.

\section{Noether charge and thermodynamics}
\label{sec:noether}

It remains to relate the relative entropy~\eqref{eq:relative_entropy_cauchy_tilted_3} to a Noether charge. Those are given by integrating the normal component of a conserved current over a Cauchy surface, and since the relative entropy is already given by an integral over a Cauchy surface, we only need to determine the corresponding conserved current. From the explicit form of $\mathcal{Q}$~\eqref{eq:q_def} as well as the fact that the energy density~\eqref{eq:energy_density} appears, we infer that the current involves the stress tensor. Because the modular Hamiltonian~\eqref{eq:modular_hamiltonian} generates translations along the trajectories~\eqref{eq:wedge_trajectory_embedding} whose tangent vector is a Killing vector, it is further probable that the current is obtained by contracting the stress tensor $T_{\mu\nu}$ with the Killing vector $- M_{0j} = \xi_{(j)}^\nu \partial_\nu$~\eqref{eq:killing_m0j}: $J^\mu \equiv T^{\mu\nu} \xi^{(j)}_\nu$. Namely, in this case we obtain
\begin{equation}
\nabla_\mu J^\mu = \nabla_{(\mu} \xi^{(j)}_{\nu)} T^{\mu\nu} + \xi^{(j)}_\nu \nabla_\mu T^{\mu\nu} = 0 \eqend{,}
\end{equation}
employing the Killing equation $\nabla_{(\mu} \xi^{(j)}_{\nu)} = 0$ (since the stress tensor is symmetric) and the conservation of the stress tensor $\nabla_\mu T^{\mu\nu} = 0$ which holds on shell.

It turns out that this indeed holds, and with the canonical scalar stress tensor
\begin{equation}
\label{eq:stress_tensor_def}
T_{\mu\nu}(h) \equiv \partial_\mu h \partial_\nu h - \frac{1}{2} g_{\mu\nu} \left( g^{\rho\sigma} \partial_\rho h \partial_\sigma h + m^2 h^2 \right)
\end{equation}
we compute on the Cauchy surface $\Sigma$~\eqref{eq:cauchy_surface_tilted_main}
\begin{equation}
\label{eq:q_noether}
n^\mu \xi_{(j)}^\nu T_{\mu\nu}(h) = \mathcal{Q}_j(h) \eqend{.}
\end{equation}
Since the commutator function $\Delta$ fulfills the Klein--Gordon equation, the classical stress tensor~\eqref{eq:stress_tensor_def} evaluated on $h = \Delta f$ is conserved, such that the current $J^\mu = \xi^{(j)}_\nu T^{\mu\nu}(\Delta f)$ is conserved as well. It follows that $\mathcal{Q}_j(\Delta f)$, the integrand of the relative entropy~\eqref{eq:relative_entropy_cauchy_tilted_3}, is its normal component, and so the relative entropy $\mathcal{S}(f \Vert \Omega)$ \emph{is} the Noether charge --- if $f$ is supported in the wedge $\mathcal{W}_j$~\eqref{eq:dswedge_poincare} such that $\Delta f$, restricted to the Cauchy surface, has support only where $\vec{x}^j \geq 0$ and $\Theta(\vec{x}^j) = 1$. Otherwise, because $\mathcal{Q}_j$ is negative for $\vec{x}^j < 0$ as can be seen from the explicit expression~\eqref{eq:q_def}, the Noether charge is strictly smaller than the relative entropy.

The definition of the relative entropy $\mathcal{S}(\rho\Vert\sigma)$~\eqref{eq:relative_entropy} leads directly to its microscopic interpretation~\cite{vedral2002}, namely as the average amount of information that is gained from measurements when the system is mistakenly described by the density matrix $\sigma$, while its true density matrix is given by $\rho$. In other words, after making a large number $N$ of measurements, the probability $p$ to mistakenly describe the system by $\sigma$ while it actually is in the state described by $\rho$ is proportional to $\mathe^{- N \mathcal{S}(\rho\Vert\sigma)}$. However, it is also possible to give a macroscopic interpretation in a thermodynamic context. Instead of the usual von Neumann entropy, also relative entropy can be used to formulate the thermodynamic laws~\cite{floerchingerhaas2020,dowlingfloerchingerhaas2020}. In particular, the second law according to which entropy can only increase in a physical process can be reformulated as the statement that relative entropy in some given spacetime volume can only \emph{decrease}, such that the states become less distinguishable, and the third law can be replaced by the statement that the relative entropy between the ground state and a thermodynamic equilibrium state (in either the canonical or grand-canonical ensembles) vanishes in the limit of vanishing temperature $T \to 0$. Instead, the first law still involves the entropy $S$ itself, and in the microcanonical ensemble reads
\begin{equation}
\label{eq:entropy_variation}
\delta \mathcal{S}\left( \rho \Vert \sigma_\text{m} \right) = - \delta S(\rho) + \beta \delta E(\rho) - \beta \mu \, \delta N(\rho) \eqend{.}
\end{equation}
Here, $\rho$ is the density matrix of an arbitrary state with the same volume, energy $E$, and particle number $N$ as the microcanonical density matrix $\sigma_\text{m}$, $\beta$ is the inverse temperature and $\mu$ the chemical potential. In the canonical and grand-canonical ensembles, a similar relation holds.

Consider now the relative entropy for two different coherent excitations with test functions $h$ and $h + \delta h$. For a coherent excitation, we have $\rho_h = U(h) \rho \, U^{-1}(h)$ with the unitary $U(h) = \mathe^{\mathi \phi(h)}$, from which it follows that $f(\rho_h) = U(h) f(\rho) U^{-1}(h)$ for reasonable functions $f$. Consequently, we have $\tr( \rho_f \ln \rho_f ) = \tr( \rho \ln \rho )$ and the von Neumann entropies $S(\rho)$ and $S(\rho_f)$ are equal. It follows that one can obtain the (inverse) temperature $\beta$ associated to the state by deriving the relative entropy of a coherent excitation with respect to the energy, while the chemical potential is obtained by deriving it with respect to the particle number. Now our result~\eqref{eq:relative_entropy_cauchy_tilted_3} is of the form
\begin{equation}
\label{eq:relative_entropy_cauchy_tilted_4}
\mathcal{S}(f \Vert \Omega) = \int \beta^\mu T_{\mu\nu}(\Delta f) n^\nu \big\rvert_\Sigma \sqrt{\gamma} \total^{n-1} \vec{x}
\end{equation}
with the local temperature vector
\begin{equation}
\label{eq:temperature_vector}
\beta^\mu = 2 \pi \Theta(\vec{x}^j) \xi_{(j)}^\mu \eqend{.}
\end{equation}
This is clearly a generalization of the global thermodynamic laws to the case where quantities such as the temperature can vary with the position. Consider then an observer who follows the trajectories~\eqref{eq:wedge_trajectory_embedding}, such that their four-velocity $u^\mu$ is proportional to the Killing vector $\xi_{(j)}^\mu$. On the Cauchy surface $\Sigma$, using Eqs.~\eqref{eq:killing_m0j} and~\eqref{eq:cauchy_surface_tilted} we compute that
\begin{equation}
u^\mu \big\rvert_\Sigma = \frac{\sqrt{ 1 + H^2 \vec{x}^2 }}{\vec{x}^j} \xi_{(j)}^\mu \big\rvert_\Sigma = n^\mu \big\rvert_\Sigma \eqend{,}
\end{equation}
which is normalized to $u^\mu u_\mu = -1$, such that the local temperature vector~\eqref{eq:temperature_vector} can be written as $\beta^\mu = \beta u^\mu$ with the local temperature
\begin{equation}
\beta(x) \big\rvert_\Sigma = 2 \pi \Theta(\vec{x}^j) \frac{\vec{x}^j}{\sqrt{ 1 + H^2 \vec{x}^2 }} \eqend{.}
\end{equation}
Indeed, with these definitions the relative entropy~\eqref{eq:relative_entropy_cauchy_tilted_4} can be written in the form
\begin{equation}
\mathcal{S}(f \Vert \Omega) = \int \beta e(\Delta f) \big\rvert_\Sigma \sqrt{\gamma} \total^{n-1} \vec{x}
\end{equation}
with the local energy density $e(\Delta f) \equiv u^\mu u^\nu T_{\mu\nu}(\Delta f)$, and the local inverse temperature $\beta$ is correctly obtained as the variation of the relative entropy with respect to the energy density that the observer sees. In the flat-space limit $H \to 0$, we moreover recover the known result $\beta(x) \big\rvert_\Sigma \to 2 \pi \Theta(\vec{x}^j) \vec{x}^j$~\cite{buchholzsolveen2013,ariasblancocasinihuerta2017}.

\section{Discussion}
\label{sec:discussion}

We have considered the relative entropy between the vacuum and a coherent excitation thereof for a massive scalar field in de Sitter spacetime. The result~\eqref{eq:relative_entropy_cauchy_tilted_3} reads
\begin{equation}
\mathcal{S}(f\Vert \Omega) = 2 \pi \int_{\vec{x}^j \geq 0} \mathcal{Q}_j(\Delta f) \big\rvert_\Sigma \sqrt{\gamma} \total^{n-1} \vec{x}
\end{equation}
with $\mathcal{Q}_j(h) = n^\mu \xi_{(j)}^\nu T_{\mu\nu}(h)$~\eqref{eq:q_noether}, and is equal to the Noether charge of translations along the flow of the Killing vector field $- M_{0j} = \xi_{(j)}^\nu \partial_\nu$ if $f$ has support in the wedge $\mathcal{W}_j$~\eqref{eq:dswedge_poincare}. While we have derived this result in all detail for the case of free scalar fields, we expect it to hold more generally because of the following reasons: by the Bisognano--Wichmann theorem~\cite{bisognanowichmann1975,bisognanowichmann1976} the modular Hamiltonian $\ln \Delta_\Omega$ for the vacuum state restricted to wedges in Minkowski spacetime is equal to $- 2 \pi \mathcal{M}_{0j}$, even for interacting fields. Since the generators $M_{0j}$ are tangent to the de~Sitter hyperboloid, also the modular Hamiltonian for de~Sitter wedges is equal to $- 2 \pi \mathcal{M}_{0j}$~\cite{borchersbuchholz1999}. Because $M_{0j}$ generates Lorentz or de~Sitter symmetries, the action of $\mathcal{M}_{0j}$ on quantum fields is given by the commutator with the stress tensor, contracted with the Killing vector $\xi_{(j)}^\nu$, and this holds also for interacting theories. Of course, what may happen is that the commutator with the Weyl operators $W(f) = \mathe^{\mathi \phi(f)}$ does not result anymore in the classical stress tensor evaluated on $\Delta f$, and this is why we cannot ascertain that our result will hold in general. It is also possible that the result in this case may depend on the renormalization freedom of the stress tensor in the quantum theory~\cite{hollandswald2005}, and it would therefore be very interesting to study the interacting case in detail, even at first order in perturbation theory.

We have further verified that the relative entropy $\mathcal{S}(f\Vert \Omega)$ is positive, for which we needed to choose a suitable Cauchy surface~\eqref{eq:cauchy_surface_tilted_main} to make positivity manifest. We have to also checked its joint convexity~\eqref{eq:relative_entropy_convex}, which for the free scalar field follows more or less straightforwardly from positivity. Since these two properties follow from general considerations for the relative entropy~\cite{araki1976,witten2018}, they provide a useful check on the computation, and of course will also hold in more general situations including interactions. Lastly, we have determined the local temperature that is seen by an observer who is following the flow trajectories of the Killing vector field $- M_{0j}$, generalizing the result for wedges in flat spacetime~\cite{buchholzsolveen2013,ariasblancocasinihuerta2017}. We have done this by employing the reformulation of thermodynamic laws using relative entropy~\cite{floerchingerhaas2020,dowlingfloerchingerhaas2020}, which for a coherent excitation tells us that the local inverse temperature can be obtained as the derivative of the relative entropy with respect to the energy density that the observer sees.

Another important property of relative entropy is its convexity with respect to inclusions~\cite{ciollilongoranalloruzzi2022}. For this, let us consider instead of a single wedge $\mathcal{W}_j$ a family of regions $\mathcal{R}(\lambda)$ depending on a parameter $\lambda$, which are such that the modular automorphism $\alpha_s$~\eqref{eq:modular_automorphism} with positive flow parameter $s$ moves operators into regions with larger $\lambda$. That is, given $W(f)$ with $\supp f \subset \mathcal{R}(\lambda_0)$, for each $\lambda \geq \lambda_0$ there exists some $s \geq 0$ such that $\supp \alpha_s(W(f)) \subset \mathcal{R}(\lambda)$. Such a construction is known as half-sided modular inclusion, and holds for example for a family of wedges in Minkowski spacetime which are translated in a null direction~\cite{ciollilongoruzzi2020,ciollilongoranalloruzzi2022,morinellitanimotowegener2022}. Computing the relative entropy $\mathcal{S}_\lambda(f\Vert \Omega)$ for this family of regions, its convexity is the statement that $\partial_\lambda^2 \mathcal{S}_\lambda(f\Vert \Omega) \geq 0$. For the null translated Minkowski wedges, it turns out that the second derivative of the relative entropy is given by (see Theorem~3.6 in Ref.~\cite{ciollilongoruzzi2020})
\begin{equation}
\label{eq:convexity_hsmi}
\partial_\lambda^2 \mathcal{S}_\lambda(f\Vert \Omega) = 2 \pi \int v^\mu v^\nu T_{\mu\nu}(\Delta f) \big\rvert_{t = \vec{x}^j = \lambda} \total^{n-2} \vec{x}^\perp \eqend{,}
\end{equation}
where $v^\mu = \delta^\mu_0 + \delta^\mu_j$ is the lightlike vector tangent to the upper null boundary of the wedge $W_j$. The pointwise condition $v^\mu v^\nu T_{\mu\nu} \geq 0$ is known as the null energy condition (NEC), and expresses the fact that light rays are always focused and never repelled by matter. While classical matter usually does obey this condition as well as other energy conditions (and in fact the NEC does hold for the case at hand~\cite{ciollilongoruzzi2020}), in the quantum theory pointwise conditions are violated, and moreover one can find states in which the energy density at any single point can have arbitrarily negative expectation values~\cite{epsteinglaserjaffe1965,fewster2006}. However, it is possible to derive averaged energy conditions (see Refs.~\cite{fewster2012,kontousanders2020} and references therein), where the pointwise condition is smeared with a test function, at least for the weak energy condition where the stress tensor is contracted with the tangent vector of a timelike worldline. While the lower bound depends on the test function and can be negative in general, at least it is finite. The situation becomes better in the limit where the test function $f$ becomes a constant, and the corresponding condition is called averaged. In particular, the averaged NEC (ANEC) very often holds, in many cases even with a lower bound of $0$~\cite{klinkhammer1991,folacci1992,verch2000,hartmankundutajdini2017}.

Clearly, the (pointwise) NEC implies convexity of the relative entropy~\eqref{eq:convexity_hsmi}, but the reverse is not necessarily true. Moreover, also the ANEC does not imply convexity, since the integral in Eq.~\eqref{eq:convexity_hsmi} is over directions transverse to the null boundary and not along it. Nevertheless, the convexity of the relative entropy is related to a NEC of a different type, dubbed the quantum NEC (QNEC), which reads $v^\mu v^\nu \bra{\Psi} T_{\mu\nu}(x) \ket{\Psi} \geq \mathcal{S}''_\Psi$. Here, $\Psi$ is a state restricted to a region whose boundary contains $x$, and $\mathcal{S}''_\Psi$ is the second derivative of the von~Neumann entropy with respect to null translations of this region along $v^\mu$. Even though the von~Neumann entropy itself is divergent, its derivative can be finite, and the QNEC has been argued to hold in general~\cite{boussoetal2015,boussoetal2016a,boussoetal2016b}. Moreover, the relation between the QNEC, the ANEC and modular theory has been elucidated~\cite{faulkneretal2016,ceyhanfaulkner2020,longo2020}.

It would of course be interesting to prove convexity for inclusions also in the de~Sitter case, and to see if one could relate it to a QNEC or ANEC. However, the wedges $\mathcal{W}_j$ in de~Sitter space are rather artificial regions, as can be seen from Fig.~\ref{fig:dswedge}. The regions which play a more physical role are the double cones or diamonds inside the static patch, which are the regions that a geodesic observer can causally influence. The modular Hamiltonian for double cones in de~Sitter has been determined recently~\cite{froeb2023}, and we plan to extend our computation also to this setting. Moreover, the diamonds can be chosen such that their (future or past) boundary coincides with the cosmological horizon, which allows us to study the generalized second law~\cite{bekenstein1973,bekenstein1974} (with the generalized entropy being given by the sum of the matter entropy and the Bekenstein--Hawking entropy of the horizon) from a rigorous field-theoretic perspective~\cite{dangeloetal2023}. In turn, we could then compare the obtained results with the ones obtained in the de~Sitter holographic correspondence~\cite{strominger2001,alishahihakarchsilversteintong2004,alishahihakarchsilverstein2005}, where the horizon horizon entropy can be understood as the entanglement entropy between the right and left dual conformal field theories (CFTs) that appear in the correspondence, or between the past and future dual CFTs~\cite{nguyen2017,narayan2018,dongsilversteintorroba2018,genggrieningerkarch2019,ariasdiazsundell2020,geng2020,geng2021}.

\begin{acknowledgments}
This work was supported and funded by \emph{Kuwait University}, Research Project No.~SM06/21.
\end{acknowledgments}

\appendix

\section{Covariant canonical quantization}
\label{sec:app_quantization}

Here we recall the covariant canonical quantization of a scalar field; see Refs.~\cite{crnkovicwitten1987,higuchi1989} for a more detailed discussion. The Lagrangian density for a massive scalar field with mass $m$ in a curved spacetime reads
\begin{splitequation}
\mathcal{L} &= - \frac{1}{2} \left( \nabla^\mu \phi \nabla_\mu \phi + m^2 \phi^2 \right) \sqrt{-g} \\
&= \frac{1}{2} \mathe^{(n-1) H t} \left[ \dot \phi^2 - \mathe^{- 2 H t} \partial^i \phi \partial_i \phi - m^2 \phi^2 \right] \eqend{,}
\end{splitequation}
where the second expression holds in the expanding Poincar{\'e} patch with metric~\eqref{eq:poincare_metric}, and we denote the derivative with respect to the cosmological time $t$ by a dot. In the following, we write $x^\mu = (t,\vec{x})$ and $y^\mu = (s,\vec{y})$. The covariant canonical momentum is defined from the Lagrangian density by
\begin{equation}
\pi^\mu \equiv \frac{1}{\sqrt{-g}} \frac{\partial \mathcal{L}}{\partial (\nabla_\mu \phi)} = - \nabla^\mu \phi \eqend{,}
\end{equation}
and the Euler--Lagrange equation for $\phi$ is the Klein--Gordon equation (with $\laplace \equiv \partial^i \partial_i$)
\begin{splitequation}
\label{eq:klein_gordon}
&\left( \nabla^2 - m^2 \right) \phi \\
&= \left[ - \partial_t^2 - (n-1) H \partial_t + \mathe^{- 2 H t} \laplace - m^2 \right] \phi = 0 \eqend{.}
\end{splitequation}
For any two solutions $\phi_{(1)}$ and $\phi_{(2)}$, it follows that the covariant symplectic current
\begin{equation}
J_{(1,2)}^\mu \equiv \mathi \left( \phi_{(1)}^* \pi_{(2)}^\mu - \phi_{(2)} \pi_{(1)}^{\mu*} \right)
\end{equation}
is covariantly conserved: $\nabla_\mu J_{(1,2)}^\mu = 0$. Consequently, the covariant symplectic product
\begin{equation}
\label{eq:symplectic_product}
\symp{ \phi_{(1)}, \phi_{(2)} } \equiv \int_\Sigma J_{(1,2)}^\mu n_\mu \total^{n-1} \Sigma \eqend{,}
\end{equation}
where $\Sigma$ is a Cauchy surface, $n_\mu$ the future-directed unit vector normal to $\Sigma$ and $\total^{n-1} \Sigma$ the normalized surface element, is independent of the choice of the Cauchy surface (assuming the solutions fall off fast at spatial infinity). This follows straightforwardly; we have
\begin{splitequation}
0 &= \int_V \nabla_\mu J_{(1,2)}^\mu \sqrt{-g} \total^n x = \int_V \partial_\mu \left( J_{(1,2)}^\mu \sqrt{-g} \right) \total^n x \\
&= \int_{\partial V} J_{(1,2)}^\mu \sqrt{-g} \, n_\mu \total^{n-1} S \raisetag{1.8em}
\end{splitequation}
for any region $V$ with boundary $\partial V$ by Gauß' divergence theorem, where $n_\mu$ is the outward-directed unit vector normal to the boundary $\partial V$ and $\total^{n-1} S$ the normalized surface element of the boundary. Choosing $V$ to be the complete volume between two Cauchy surfaces $\Sigma_1$ and $\Sigma_2$ such that $\partial V = \Sigma_1 \cup \Sigma_2 \cup S_\infty$, where $S_\infty$ is the part of the boundary at spatial infinity, assuming that the solutions fall off fast at spatial infinity such that $J_{(1,2)}^\mu$ vanishes there, and flipping the normal vector of the earlier Cauchy surface $\Sigma_1$ such it is also future-pointing, we obtain
\begin{equation}
0 = \int_{\Sigma_1} J_{(1,2)}^\mu n_\mu \total \Sigma - \int_{\Sigma_2} J_{(1,2)}^\mu n_\mu \total^{n-1} \Sigma
\end{equation}
with $\total^{n-1} \Sigma = \sqrt{-g} \total^{n-1} S$, and thus the independence of the covariant symplectic product from the Cauchy surface. For our purposes, we can choose the Cauchy surface $\Sigma$ to be at constant time $t = 0$ such that $n_\mu = \delta_\mu^0$ and $\sqrt{-g} = 1$, such that
\begin{splitequation}
\label{eq:symplectic_product_poincare}
&\symp{ \phi_{(1)}, \phi_{(2)} } \\
&= \mathi \int \left[ \phi_{(1)}^*(0,\vec{x}) \dot \phi_{(2)}(0,\vec{x}) - \phi_{(2)}(0,\vec{x}) \dot \phi_{(1)}^*(0,\vec{x}) \right] \total^{n-1} \vec{x} \eqend{,} \raisetag{3.8em}
\end{splitequation}
where we recall that a dot denotes a derivative with respect to time. In the embedding coordinates~\eqref{eq:poincarepatch}, this corresponds to the surface $X^0 - X^n = H^{-1}$, and also the antisymmetry
\begin{equation}
\label{eq:symplectic_product_antisymmetry}
\symp{ \phi_{(2)}, \phi_{(1)} } = - \symp{ \phi_{(1)}^*, \phi_{(2)}^* }
\end{equation}
is manifest.

Since the symplectic product is nondegenerate, we obtain a complete set of normalizable mode solutions $f_\vec{p}(t,\vec{x})$ to the Klein--Gordon equation,
\begin{splitequation}
\label{eq:mode_functions_bd}
f_\vec{p}(t,\vec{x}) &= c(\vec{p}) \, \mathe^{- \frac{n-1}{2} H t} \, \mathe^{\mathi \vec{p} \vec{x}} \\
&\quad\times \left[ \bessel{J}{\mu}{\frac{\abs{\vec{p}}}{H} \mathe^{- H t}} + \mathi \, \bessel{Y}{\mu}{\frac{\abs{\vec{p}}}{H} \mathe^{- H t}} \right]
\end{splitequation}
where $\mu \equiv \sqrt{ \frac{(n-1)^2}{4} - \frac{m^2}{H^2} }$, $\mathrm{J}$ and $\mathrm{Y}$ are the Bessel functions of the first and second kind, and the factor $c(\vec{p})$ can be determined from the normalization. Note that for small masses $m^2 \leq (n-1)^2/4 \, H^2$, $\mu$ is real, but for large masses it becomes purely imaginary: in this case, one has to use the appropriate real and imaginary parts $\tilde{\mathrm{J}}$ and $\tilde{\mathrm{Y}}$~\cite{dlmf}. In the following, we consider only real $\mu$, and leave the straightforward modifications for purely imaginary $\mu$ to the reader. Since for real argument and parameter the Bessel functions are real~\cite{dlmf}, we obtain
\begin{splitequation}
f^*_\vec{p}(t,\vec{x}) &= c^*(\vec{p}) \, \mathe^{- \frac{n-1}{2} H t} \, \mathe^{- \mathi \vec{p} \vec{x}} \\
&\quad\times \left[ \bessel{J}{\mu}{\frac{\abs{\vec{p}}}{H} \mathe^{- H t}} - \mathi \, \bessel{Y}{\mu}{\frac{\abs{\vec{p}}}{H} \mathe^{- H t}} \right] \eqend{,}
\end{splitequation}
and using Bessel function identities, for the symplectic product~\eqref{eq:symplectic_product} of two modes\footnote{Even though the modes do not decay at spatial infinity, one can nevertheless easily see explicitly that their symplectic product is time-independent by reinstating a factor $\sqrt{-g} = \mathe^{(n-1) H t}$ in the covariant symplectic product~\eqref{eq:symplectic_product} and evaluating it at a generic time $t$.} we obtain
\begin{equation}
\label{eq:sympletic_product_modefunctions1}
\symp{ f_\vec{p}, f_\vec{q} } = (2\pi)^{n-1} \delta^{n-1}(\vec{p}-\vec{q}) \abs{ c(\vec{p}) }^2 \frac{4 H}{\pi} \eqend{.}
\end{equation}
We thus choose $\abs{ c(\vec{p}) } = \sqrt{\pi/(4H)}$, and leave the phase factor undetermined, since it will not affect the result.\footnote{To obtain the correct flat-space limit of the modes, one has to choose the phase factor as in Ref.~\cite{nachtmann1967}.} Similarly, we obtain
\begin{equations}[eq:sympletic_product_modefunctions2]
\symp{ f^*_\vec{p}, f_\vec{q} } &= 0 = \symp{ f_\vec{p}, f^*_\vec{q} } \eqend{,} \\
\symp{ f^*_\vec{p}, f^*_\vec{q} } &= (2\pi)^{n-1} \delta^{n-1}(\vec{p}-\vec{q}) \eqend{,}
\end{equations}
which reflects the orthogonality of the set of mode functions. Moreover, we have a second orthogonality relation in Fourier space, which reads
\begin{splitequation}
\label{eq:mode_function_orthogonal2}
&\mathi \int \left[ f^*_\vec{p}(t,\vec{x}) \dot f_\vec{p}(t,\vec{y}) - f_\vec{p}(t,\vec{x}) \dot f^*_\vec{p}(t,\vec{y}) \right] \frac{\total^{n-1} \vec{p}}{(2\pi)^{n-1}}\\
&\quad= \mathe^{-(n-1) H t} \delta^{n-1}(\vec{x}-\vec{y}) \eqend{.}
\end{splitequation} 

The field is then canonically quantized as
\begin{splitequation}
\label{eq:phi_modes}
\phi(x) &= \int \left[ a(\vec{p}) f_\vec{p}(t,\vec{x}) + a^\dagger(\vec{p}) f^*_\vec{p}(t,\vec{x}) \right] \frac{\total^{n-1} \vec{p}}{(2\pi)^{n-1}} \\
&= \phi^\dagger(x) \eqend{,} \raisetag{1.4em}
\end{splitequation}
where the creation and annihilation operators fulfill the commutation relations
\begin{equations}[]
[ a(\vec{p}), a^\dagger(\vec{q}) ] &= \symp{ f_\vec{p}, f_\vec{q} } = (2\pi)^{n-1} \delta^{n-1}(\vec{p}-\vec{q}) \eqend{,} \\
[ a(\vec{p}), a(\vec{q}) ] &= 0 = [ a^\dagger(\vec{p}), a^\dagger(\vec{q}) ] \eqend{,}
\end{equations}
and can be obtained from the field by computing the sympletic product with the mode functions:
\begin{equation}
\label{eq:a_adagger_symplectic}
\symp{ f_\vec{p}, \phi } = a(\vec{p}) \eqend{,} \quad \symp{ f^*_\vec{p}, \phi } = a^\dagger(\vec{p}) \eqend{.}
\end{equation}
The vacuum vector $\ket{\Omega}$ is annihilated by all annihilation operators $a(\vec{p})$, and for the above choice of mode functions is known as the Bunch--Davies vacuum~\cite{chernikovtagirov1968,schomblondspinde1976,bunchdavies1978}. Using the second orthogonality relation, it follows that $\phi$ and $n_\mu \pi^\mu = \dot \phi$ fulfill the covariant canonical commutation relations on the Cauchy surface $\Sigma = \{ (t,\vec{x})\colon t = \text{const} \}$:
\begin{equation}
\label{eq:commutator_phi_canonical}
[ \phi(x), \dot \phi(y) ] \Big\rvert_\Sigma = \mathi \mathe^{-(n-1) H t} \delta^{n-1}(\vec{x}-\vec{y}) = \mathi \frac{\delta^{n-1}(\vec{x}-\vec{y})}{\sqrt{-g}} \eqend{.}
\end{equation}

From the expansion of $\phi$ in modes~\eqref{eq:phi_modes} and the explicit form of the modes~\eqref{eq:mode_functions_bd}, we can now compute the Wightman two-point function
\begin{equation}
\label{eq:wightman_def}
\bra{\Omega} \phi(x) \phi(x') \ket{\Omega} = \int f_\vec{p}(x) f^*_\vec{p}(x') \frac{\total^{n-1} \vec{p}}{(2\pi)^{n-1}} \eqend{.}
\end{equation}
The computation of the integral is quite involved, and is done in detail in Appendix A in Ref.~\cite{froebhiguchi2014}. It results in
\begin{equation}
\label{eq:wightman_2pf}
\bra{\Omega} \phi(x) \phi(x') \ket{\Omega} = \frac{H^{n-2}}{(4\pi)^\frac{n}{2}} I_\mu(x,x') \eqend{,}
\end{equation}
with the function
\begin{widetext}
\begin{equation}
\label{eq:imu_integral}
I_\mu(x,x') = \frac{\Gamma\left( \frac{n-1}{2}+\mu \right) \Gamma\left( \frac{n-1}{2}-\mu \right)}{\Gamma\left( \frac{n}{2} \right)} \lim_{\epsilon \to 0^+} \hypergeom{2}{1}\left[ \frac{n-1}{2}+\mu, \frac{n-1}{2}-\mu; \frac{n}{2}; \frac{1+Z(x,x')}{2} - \mathi \epsilon \sgn(t-t') \right]
\end{equation}
depending on the de~Sitter-invariant distance $Z(x,x')$~\eqref{eq:z_def}. For spacelike separation with $Z(x,x') < 1$, we can take the limit $\epsilon \to 0$ inside the hypergeometric function, while for timelike separation with $Z(x,x') > 1$ we first have to use a transformation of variables [see Eq.~(15.8.3) in Ref.~\cite{dlmf}] to obtain
\begin{splitequation}
\label{eq:imu_integral_timelike}
I_\mu(x,x') &= \frac{2^{\frac{n-3}{2}-\mu} \Gamma\left( \frac{n-1}{2} + \mu \right) \Gamma(-\mu)}{\sqrt{\pi}} \left[ Z(x,x') - 1 \right]^{- \frac{n-1}{2} - \mu} \mathe^{- \mathi \pi \left( \frac{n-1}{2}+\mu \right) \sgn(t-t') } \\
&\quad\times \hypergeom{2}{1}\left( \frac{1}{2} + \mu, \frac{n-1}{2}+\mu ; 1+2\mu; - \frac{2}{Z(x,x')-1} \right) + (\mu \rightarrow -\mu) \eqend{.}
\end{splitequation}
\end{widetext}

It is also useful to write the result~\eqref{eq:imu_integral} in a slightly different form, taking into account that the causal structure of de~Sitter space is inherited from the ambient space, such that $\sgn(t-t') = \sgn\left( X^0(x)-X^0(x') \right)$ if the points are timelike separated. Using the expression of $Z$ in terms of embedding coordinates~\eqref{eq:z_embedding} we obtain
\begin{equation}
\lim_{\epsilon \to 0^+} \left[ \frac{1+Z(x,x')}{2} - \mathi \epsilon \sgn(t-t') \right] = \lim_{\epsilon \to 0^+} \frac{1+Z_\epsilon(x,x')}{2}
\end{equation}
with
\begin{splitequation}
\label{eq:zepsilon_def}
Z_\epsilon(x,x') &= - H^2 \big[ X^0(x) - \mathi \epsilon \big] \big[ X^0(x') + \mathi \epsilon \big] \\
&\quad+ H^2 X_i(x) X^i(x') + H^2 X^n(x) X^n(x') \eqend{,} \raisetag{2.9em}
\end{splitequation}
since in the limit $\epsilon \to 0$ only the sign is relevant.

\section{Commutator and symplectic product}
\label{sec:app_symplectic}

From the mode functions~\eqref{eq:mode_functions_bd}, one also computes the advanced and retarded propagators
\begin{equations}[eq:green_advanced_retarded]
\begin{split}
G_\text{ret}(x,y) &= - \mathi \Theta(t-s) \int \Big[ f_\vec{p}(t,\vec{x}) f^*_\vec{p}(s,\vec{y}) \\
&\qquad\quad- f^*_\vec{p}(t,\vec{x}) f_\vec{p}(s,\vec{y}) \Big] \frac{\total^{n-1} \vec{p}}{(2\pi)^{n-1}} \eqend{,}
\end{split} \\
G_\text{adv}(x,y) &= G_\text{ret}(y,x) \eqend{,}
\end{equations}
which are easily checked to be fundamental solutions of the Klein--Gordon equation, using again the second orthogonality relation:
\begin{splitequation}
&\left[ - \partial_t^2 - (n-1) H \partial_t + \mathe^{- 2 H t} \laplace - m^2 \right] G_\text{ret/adv}(x,y) \\
&\quad= \mathe^{- (n-1) H t} \delta(t-s) \delta^{n-1}(\vec{x}-\vec{y}) = \frac{\delta^n(x-y)}{\sqrt{-g}} \eqend{.}
\end{splitequation}
Finally, the commutator function
\begin{splitequation}
\label{eq:commutator_delta}
&\Delta(x,y) \equiv G_\text{adv}(x,y) - G_\text{ret}(x,y) \\
&= \mathi \int \left[ f_\vec{p}(t,\vec{x}) f^*_\vec{p}(s,\vec{y}) - f^*_\vec{p}(t,\vec{x}) f_\vec{p}(s,\vec{y}) \right] \frac{\total^{n-1} \vec{p}}{(2\pi)^{n-1}} \eqend{,}
\end{splitequation}
which satisfies the Klein--Gordon equation~\eqref{eq:klein_gordon}, determines the spacetime commutator of two fields:
\begin{equation}
\label{eq:commutator_phi}
[ \phi(x), \phi(y) ] = - \mathi \Delta(x,y) \eqend{.}
\end{equation}
For the commutator function we have the important identity $\Delta(t,\vec{x},t,\vec{y}) = 0$, which is proven by changing $\vec{p} \to - \vec{p}$ in the second term of Eq.~\eqref{eq:commutator_delta}, and reflects the fact that the fields commute at spatial separations. Analogously, we obtain $\partial_s \partial_t \Delta(x,y) \big\rvert_{s = t} = 0$, which encodes the vanishing of the canonical momenta at spatial separations. Furthermore, using the second orthogonality relation~\eqref{eq:mode_function_orthogonal2}, it also holds that
\begin{equation}
\partial_t \Delta(x,y) \Big\rvert_{s = t} = \mathe^{-(n-1) H t} \delta^{n-1}(\vec{x}-\vec{y}) \eqend{,}
\end{equation}
which reflects that the field and momentum have the non-vanishing equal-time commutator~\eqref{eq:commutator_phi_canonical}.

In general, we consider fields smeared with a test function $f$
\begin{equation}
\phi(f) = \int f(x) \phi(x) \sqrt{-g} \total^n x \eqend{,}
\end{equation}
where $f \in \mathcal{S}(\mathbb{R}^n)$ is a Schwartz function, and for two test functions define an inner product
\begin{equation}
\left( f, h \right) \equiv \int f^*(x) h(x) \sqrt{-g(x)} \total^n x \eqend{.}
\end{equation}
It follows that the commutator~\eqref{eq:commutator_phi} can be written in the form
\begin{equation}
\label{eq:commutator_phi_smeared}
[ \phi(f), \phi(h) ] = - \mathi \left( f^*, \Delta h \right) = - \mathi \left( \Delta f^*, h \right) \eqend{,}
\end{equation}
where
\begin{equation}
\label{eq:delta_smeared_def}
(\Delta h)(x) \equiv \int \Delta(x,y) h(y) \sqrt{-g(y)} \total^n y
\end{equation}
is the smeared commutator function.

Consider now two functions $f$ and $h$, where at least one of them has compact support in spatial directions. Then Green's identity (i.e., integration by parts) results in
\begin{widetext}
\begin{splitequation}
&\int_{t_1 \leq x^0 \leq t_2} \left[ f^*(x) \left( \nabla^2 - m^2 \right) h(x) - h(x) \left( \nabla^2 - m^2 \right) f^*(x) \right] \sqrt{-g(x)} \total^n x \\
&\quad= - \int \left[ \mathe^{(n-1) H t} f^*(t,\vec{x}) \dot h(t,\vec{x}) - \mathe^{(n-1) H t} h(t,\vec{x}) \dot f^*(t,\vec{x}) \right]_{t_1}^{t_2} \total^{n-1} \vec{x} \eqend{.}
\end{splitequation}
\end{widetext}
We take $f$ to be a solution of the Klein--Gordon equation, and instead of $h$ the two combinations
\begin{equation}
h_\text{ret/adv}(x) = \int G_\text{ret/adv}(x,y) h(y) \sqrt{-g(y)} \total^n y \eqend{,}
\end{equation}
where $h$ has compact support in spatial directions. Since the retarded and advanced Green's functions~\eqref{eq:green_advanced_retarded} have support only inside the light cone, for all $t_1 \leq t \leq t_2$ also $h_\text{ret/adv}$ has compact support in spatial directions. It follows that
\begin{splitequation}
&\int_{t_1 \leq t \leq t_2} f^*(x) h(x) \sqrt{-g(x)} \total^n x \\
&= - \int \Big[ \mathe^{(n-1) H t} f^*(t,\vec{x}) \dot h_\text{ret/adv}(t,\vec{x}) \\
&\qquad\quad- \mathe^{(n-1) H t} h_\text{ret/adv}(t,\vec{x}) \dot f^*(t,\vec{x}) \Big]_{t_1}^{t_2} \total^{n-1} \vec{x} \eqend{.}
\end{splitequation}
Since in the limit $t \to -\infty$, the retarded combination $h_\text{ret}(x)$ vanishes, while for $t \to \infty$ the advanced combination $h_\text{adv}(x)$ vanishes, we can take $t_1 = - \infty$, $t_2 = 0$ for $h_\text{ret}$ and $t_1 = 0$, $t_2 = \infty$ for $h_\text{adv}$, and obtain
\begin{equations}[]
\begin{split}
&\int_{t \leq 0} f^*(x) h(x) \sqrt{-g(x)} \total^n x \\
&= - \int \Big[ f^*(0,\vec{x}) \dot h_\text{ret}(0,\vec{x}) - h_\text{ret}(0,\vec{x}) \dot f^*(0,\vec{x}) \Big] \total^{n-1} \vec{x} \eqend{,}
\end{split} \raisetag{5.9em} \\
\begin{split}
&\int_{0 \leq t} f^*(x) h(x) \sqrt{-g(x)} \total^n x \\
&= \int \Big[ f^*(0,\vec{x}) \dot h_\text{adv}(0,\vec{x}) - h_\text{adv}(0,\vec{x}) \dot f^*(0,\vec{x}) \Big] \total^{n-1} \vec{x} \eqend{.}
\end{split} \raisetag{4.7em}
\end{equations}
Combining both and using the definition of the commutator function~\eqref{eq:commutator_delta}, it follows that
\begin{splitequation}
\label{eq:product_symplectic}
\left( f, h \right) &= \int f^*(x) h(x) \sqrt{-g(x)} \total^n x \\
&= \int \Big[ f^*(t,\vec{x}) \partial_t (\Delta h)(t,\vec{x}) \\
&\qquad\quad- (\Delta h)(t,\vec{x}) \dot f^*(t,\vec{x}) \Big]_{t=0} \total^{n-1} \vec{x} \\
&= - \mathi \symp{ f, \Delta h } \eqend{,}
\end{splitequation}
since both $f$ (by assumption) and $\Delta h$ are solutions of the Klein--Gordon equation, such that their symplectic product $\symp{ \cdot, \cdot }$ [given in Eq.~\eqref{eq:symplectic_product_poincare}] is well-defined and time independent. Taking into account that $\phi$ is self-adjoint, we thus obtain
\begin{equation}
\label{eq:phi_smeared_symplectic}
\phi(f) = \left( \phi, f \right) = - \mathi \symp{ \phi, \Delta f } \eqend{,}
\end{equation}
which expresses the smeared field in terms of its initial data on a Cauchy surface. Moreover, it also follows that the commutator of two smeared fields~\eqref{eq:commutator_phi_smeared} can be expressed using the symplectic product:
\begin{equation}
\label{eq:commutator_phi_symplectic}
[ \phi(f), \phi(h) ] = - \mathi \left( \Delta f^*, h \right) = - \symp{ \Delta f^*, \Delta h } \eqend{.}
\end{equation}

Furthermore, using the definition of the symplectic product~\eqref{eq:symplectic_product_poincare} we obtain
\begin{splitequation}
\symp{ \phi, f } = \mathi \int \left[ \phi(0,\vec{x}) \dot f(0,\vec{x}) - f(0,\vec{x}) \pi(0,\vec{x}) \right] \total^{n-1} \vec{x} \eqend{,}
\end{splitequation}
and hence using the canonical commutation relation~\eqref{eq:commutator_phi_canonical}
\begin{equation}
\label{eq:commutator_symplectic}
[ \symp{ \phi, f }, \symp{ \phi, g } ] = - \symp{ f^*, g } \eqend{.}
\end{equation}

Lastly, we note that the de Sitter Killing vectors define (after multiplication with $\mathi$) operators that are symmetric with respect to the inner product; for test functions $f$ and $h$ we compute
\begin{splitequation}
&\left( D^\dagger f, h \right) = \left( f, D h \right) \\
&= \iint f^*(t,\vec{x}) \left( - H^{-1} \partial_t + \vec{x}^i \partial_i \right) h(t,\vec{x}) \mathe^{(n-1) H t} \total^n x \\
&= - \iint D f^*(t,\vec{x})h(t,\vec{x}) \mathe^{(n-1) H t} \total^n x \\
&= - \left( D f, h \right) \eqend{,} \raisetag{1.4em}
\end{splitequation}
and analogously
\begin{equation}
\label{eq:killing_inner_product}
\left( f, K_j h \right) = - \left( K_j f, h \right) \eqend{,} \quad \left( f, P_j h \right) = - \left( P_j f, h \right) \eqend{,}
\end{equation}
such that $\mathi D$, $\mathi K_j$, and $\mathi P_j$ are symmetric: $(\mathi D)^\dagger = \mathi D$ etc. Since the embedding space Killing vectors are constant linear combinations of these, also they are symmetric with respect to the inner product. Using the Noether method, one can then construct symmetric operators $\mathcal{D}$, $\mathcal{K}_j$, and $\mathcal{P}_j$ on the Fock space of the free scalar field, representing them as commutators:
\begin{splitequation}
\label{eq:killing_op_commutator}
&\mathi [ \mathcal{D}, \phi(x) ] = D \phi(x) \eqend{,} \quad \mathi [ \mathcal{K}_j, \phi(x) ] = K_j \phi(x) \eqend{,} \\
&\mathi [ \mathcal{P}_j, \phi(x) ] = P_j \phi(x) \eqend{.}
\end{splitequation}
For the proof that these operators are in fact self-adjoint on Fock space, we refer the reader to Ref.~\cite{borchersbuchholz1999}.

\section{Covariance of the commutator}
\label{sec:app_commutator}

Since the commutator function $\Delta(x,y)$ is de~Sitter invariant, subjecting both $x$ and $y$ to the same transformation it stays invariant. Let us show this explicitly for the generators of translations $P_j$ and boosts $K_j$~\eqref{eq:desitter_killing}. For translations, this is very simple: with the explicit expression~\eqref{eq:commutator_delta} of $\Delta$ in terms of the modes~\eqref{eq:mode_functions_bd}, we compute
\begin{widetext}
\begin{splitequation}
\label{eq:delta_covariance_p}
\left( P_j^x + P_j^y \right) \Delta(x,y) &= \mathi \int \left[ \partial_j f_\vec{p}(t,\vec{x}) f^*_\vec{p}(s,\vec{y}) - \partial_j f^*_\vec{p}(t,\vec{x}) f_\vec{p}(s,\vec{y}) + f_\vec{p}(t,\vec{x}) \partial_j f^*_\vec{p}(s,\vec{y}) - f^*_\vec{p}(t,\vec{x}) \partial_j f_\vec{p}(s,\vec{y}) \right] \frac{\total^{n-1} \vec{p}}{(2\pi)^{n-1}} \\
&= \mathi \int \left[ \mathi \vec{p}_j f_\vec{p}(t,\vec{x}) f^*_\vec{p}(s,\vec{y}) + \mathi \vec{p}_j f^*_\vec{p}(t,\vec{x}) f_\vec{p}(s,\vec{y}) - f_\vec{p}(t,\vec{x}) \mathi \vec{p}_j f^*_\vec{p}(s,\vec{y}) - f^*_\vec{p}(t,\vec{x}) \mathi \vec{p}_j f_\vec{p}(s,\vec{y}) \right] \frac{\total^{n-1} \vec{p}}{(2\pi)^{n-1}} \\
&= 0 \eqend{,} \raisetag{1.2em}
\end{splitequation}
where we denote the point on which the generators act by a superscript. For the boosts, the computation is much more involved, and we have to use the identities
\begin{equations}
\vec{x}^i f_\vec{p}(t,\vec{x}) &= - \mathi \partial_{\vec{p}^i} f_\vec{p}(t,\vec{x}) - \mathi \frac{\vec{p}^i}{\vec{p}^2} \left[ H^{-1} \partial_t f_\vec{p}(t,\vec{x}) + \frac{n-1}{2} f_\vec{p}(t,\vec{x}) \right] \eqend{,} \\
\begin{split}
\vec{p}_i \vec{x}^i \vec{x}^j f_\vec{p}(t,\vec{x}) &= - \vec{p}^i \partial_{\vec{p}^i} \partial_{\vec{p}^j} f_\vec{p}(t,\vec{x}) + \vec{p}^j H^{-2} \mathe^{- 2 H t} f_\vec{p}(t,\vec{x}) - H^{-1} \partial_{\vec{p}^j} \partial_t f_\vec{p}(t,\vec{x}) - H^{-1} \frac{\vec{p}^j}{\vec{p}^2} \vec{p}^i \partial_{\vec{p}^i} \partial_t f_\vec{p}(t,\vec{x}) \\
&\quad- \frac{n-1}{2} \partial_{\vec{p}^j} f_\vec{p}(t,\vec{x}) - \frac{n-1}{2} \frac{\vec{p}^j}{\vec{p}^2} \vec{p}^i \partial_{\vec{p}^i} f_\vec{p}(t,\vec{x}) + \frac{\vec{p}^j}{H \vec{p}^2} \partial_t f_\vec{p}(t,\vec{x}) \\
&\quad- \frac{(n-1) (n-3)}{4} \frac{\vec{p}^j}{\vec{p}^2} f_\vec{p}(t,\vec{x}) + \frac{m^2 \vec{p}^j}{H^2 \vec{p}^2} f_\vec{p}(t,\vec{x}) \eqend{,}
\end{split} \\
\begin{split}
\vec{x}^2 f_\vec{p}(t,\vec{x}) &= - \laplace_{\vec{p}} f_\vec{p}(t,\vec{x}) + H^{-2} \mathe^{- 2 H t} f_\vec{p}(t,\vec{x}) - \frac{2 \vec{p}^j}{H \vec{p}^2} \partial_{\vec{p}^j} \partial_t f_\vec{p}(t,\vec{x}) - (n-1) \frac{\vec{p}^j}{\vec{p}^2} \partial_{\vec{p}^j} f_\vec{p}(t,\vec{x}) \\
&\quad- \frac{n-3}{H \vec{p}^2} \partial_t f_\vec{p}(t,\vec{x}) - \frac{(n-1) (3n-7)}{4 \vec{p}^2} f_\vec{p}(t,\vec{x}) + \frac{m^2}{H^2 \vec{p}^2} f_\vec{p}(t,\vec{x})
\end{split}
\end{equations}
and their complex conjugates, and integrate the derivatives with respect to $\vec{p}$ by parts to obtain
\begin{equation}
\label{eq:delta_covariance_k}
\left( K_j^x + K_j^y \right) \Delta(x,y) = 0 \eqend{.}
\end{equation}
\end{widetext}

\section{Tilted Cauchy surface}
\label{sec:app_cauchy}

We consider the surface
\begin{equation}
\label{eq:cauchy_surface_tilted}
\Sigma = \left\{ (t,\vec{x})\colon 2 H t + \ln\left( 1 + H^2 \vec{x}^2 \right) = 0 \right\} \eqend{,}
\end{equation}
whose normalized future-pointing normal vector reads
\begin{equations}
n_\mu &= - N \left( 1 + H^2 \vec{x}^2, H \vec{x}_i \right)_\mu \eqend{,} \\
n^\mu &= N \left( 1 + H^2 \vec{x}^2, - \mathe^{-2 H t} H \vec{x}^i \right)^\mu \eqend{,} \\
N &\equiv \left[ ( 1 + H^2 \vec{x}^2 )^2 - \mathe^{- 2 H t} H^2 \vec{x}^2 \right]^{- \frac{1}{2}} \eqend{,}
\end{equations}
that on the surface itself reduces to
\begin{equations}[eq:normal_vector_tilted_surface]
n_\mu \big\rvert_\Sigma &= - \frac{1}{\sqrt{1 + H^2 \vec{x}^2}} \left( 1 + H^2 \vec{x}^2, H \vec{x}_i \right)_\mu \eqend{,} \\
n^\mu \big\rvert_\Sigma &= \sqrt{1 + H^2 \vec{x}^2} \left( 1, - H \vec{x}^i \right)^\mu \eqend{,} \\
N \big\rvert_\Sigma &= \frac{1}{\sqrt{1 + H^2 \vec{x}^2}} \eqend{.}
\end{equations}
The induced metric $\gamma_{ij}$ can be obtained from
\begin{equation}
\total s^2 \big\rvert_\Sigma = \gamma_{ij} \total \vec{x}^i \total \vec{x}^j
\end{equation}
and reads
\begin{equations}
\label{eq:induced_metric_tilted}
\gamma_{ij} &= \frac{\delta_{ij}}{1 + H^2 \vec{x}^2} - \frac{H^2 \vec{x}_i \vec{x}_j}{( 1 + H^2 \vec{x}^2 )^2} \eqend{,} \\
\gamma^{ij} &= (1 + H^2 \vec{x}^2) \delta^{ij} + H^2 ( 1 + H^2 \vec{x}^2 ) \vec{x}^i \vec{x}^j \eqend{,} \\
\sqrt{\gamma} &= (1 + H^2 \vec{x}^2)^{-\frac{n}{2}} \eqend{.}
\end{equations}

All derivatives can be decomposed into a derivative normal to the surface $\partial_n \equiv n^\mu \partial_\mu$, and tangential ones $\hat{\partial}_\mu \equiv ( \delta_\mu^\nu + n_\mu n^\nu ) \partial_\nu$, such that $\partial_\mu = \hat\partial_\mu - n_\mu \partial_n$, and we compute
\begin{equations}[eq:derivatives_tilted]
\partial_n &= N ( 1 + H^2 \vec{x}^2 ) \partial_t - \mathe^{-2 H t} N H \vec{x}^i \partial_i \eqend{,} \\
\begin{split}
\hat{\partial}_t &= \partial_t - N ( 1 + H^2 \vec{x}^2 ) \partial_n \\
&= - \mathe^{- 2 H t} N^2 \left[ H^2 \vec{x}^2 \partial_t - ( 1 + H^2 \vec{x}^2 ) H \vec{x}^k \partial_k \right] \eqend{,}
\end{split} \\
\begin{split}
\hat{\partial}_i &= \partial_i - N H \vec{x}^i \partial_n \\
&= \partial_i + \mathe^{- 2 H t} N^2 H^2 \vec{x}^i \vec{x}^k \partial_k - N^2 H ( 1 + H^2 \vec{x}^2 ) \vec{x}^i \partial_t
\end{split}
\end{equations}
and on the surface itself
\begin{equations}[eq:derivatives_tilted_sigma]
\partial_n \big\rvert_\Sigma &= \sqrt{1 + H^2 \vec{x}^2} \left( \partial_t - H \vec{x}^k \partial_k \right) \eqend{,} \\
\hat{\partial}_t \big\rvert_\Sigma &= - H^2 \vec{x}^2 \partial_t + ( 1 + H^2 \vec{x}^2 ) H \vec{x}^k \partial_k \eqend{,} \\
\hat{\partial}_i \big\rvert_\Sigma &= \partial_i + H^2 \vec{x}^i \vec{x}^k \partial_k - H \vec{x}^i \partial_t \eqend{.}
\end{equations}

We can then use the relation
\begin{equation}
\hat{\partial}_t = \mathe^{- 2 H t} \frac{H \vec{x}^i}{1 + H^2 \vec{x}^2} \hat{\partial}_i
\end{equation}
to replace $\hat{\partial}_t$ in all expressions and keep only $\partial_n$ and $\hat{\partial}_j$. The Klein--Gordon equation~\eqref{eq:klein_gordon} then reduces to
\begin{splitequation}
\label{eq:klein_gordon_tilted}
\left( \nabla^2 - m^2 \right) \phi \big\rvert_\Sigma &= - \partial_n^2 \phi - (n-3) H^2 \vec{x}^k \hat{\partial}_k \phi - m^2 \phi \\
&+ \left[ ( 1 + H^2 \vec{x}^2 ) \delta^{kl} - H^2 \vec{x}^k \vec{x}^l \right] \hat\partial_k \hat\partial_l \phi \eqend{,}
\end{splitequation}
and the symplectic product~\eqref{eq:symplectic_product} on $\Sigma$ reads
\begin{splitequation}
\label{eq:symplectic_product_tilted}
\symp{ \phi_{(1)}, \phi_{(2)} } &= \mathi \int \Big[ \phi_{(1)}^*(x) \partial_n \phi_{(2)}(x) \\
&\quad- \phi_{(2)}(x) \partial_n \phi_{(1)}^*(x) \Big]_\Sigma \sqrt{\gamma} \total^{n-1} \vec{x} \eqend{.}
\end{splitequation}

\begin{widetext}
For the Killing vector $M_{0j}$~\eqref{eq:killing_m0j}, using Eqs.~\eqref{eq:derivatives_tilted} we compute
\begin{splitequation}
\label{eq:killing_m0j_tilted}
M_{0j} &= - \frac{N}{2} \vec{x}^j \left( 3 + H^2 \vec{x}^2 - \mathe^{- 2 H t} \right) \partial_n + \left( 1 + H^2 \vec{x}^2 - \mathe^{- 2 H t} \right) \left( \frac{H \vec{x}^j \vec{x}^k}{1 + H^2 \vec{x}^2} \hat\partial_k - \frac{1}{2 H} \hat\partial_j \right) \eqend{,}
\end{splitequation}
and from this the commutator
\begin{splitequation}
\label{eq:m0j_commutator_tilted}
\big[ M_{0j} h \partial_n h - h \partial_n M_{0j} h \big]_\Sigma &= - \frac{\vec{x}^j}{\sqrt{ 1 + H^2 \vec{x}^2 }} \bigg[ \partial_n h \partial_n h + \big[ ( 1 + H^2 \vec{x}^2 ) \delta^{kl} - H^2 \vec{x}^k \vec{x}^l \big] \hat\partial_k h \hat\partial_l h + m^2 h^2 + h \left( \nabla^2 - m^2 \right) h \bigg]_\Sigma \\
&\quad- \left[ \hat\partial_k \left[ \frac{\vec{x}^j}{\sqrt{ 1 + H^2 \vec{x}^2 }} H^2 \vec{x}^k \vec{x}^l h \hat\partial_l h - \vec{x}^j \sqrt{ 1 + H^2 \vec{x}^2 } h \hat\partial_k h \right] \right]_\Sigma \eqend{.} \raisetag{1.8em}
\end{splitequation}
The last term is a boundary term inside the surface $\Sigma$ and thus vanishes after integration over the Cauchy surface. This can be seen also explicitly by computing
\begin{splitequation}
\label{eq:boundary_tilted}
\int \hat{\partial}_k f(x) \big\rvert_\Sigma \sqrt{\gamma} \total^{n-1} \vec{x} &= \int \hat{\partial}_k f(x) \, \mathe^{n H t} \, 2 H \delta\left[ 2 H t + \ln\left( 1 + H^2 \vec{x}^2 \right) \right] \total^n x \\
&= - 2 H \int \mathe^{n H t} f(x) \left[ \partial_k + H^2 \vec{x}^k \vec{x}^j \partial_j - H \vec{x}^k \partial_t \right] \delta\left[ 2 H t + \ln\left( 1 + H^2 \vec{x}^2 \right) \right] \total^n x = 0 \eqend{,}
\end{splitequation}
where the last equality follows after performing the derivatives on the Dirac $\delta$.
\end{widetext}

\bibliography{literature}

\end{document}